\documentclass[a4paper,10pt,twoside]{cpc-hepnp}
\usepackage{CJK,upgreek,fancyhdr}
\usepackage{multicol}
\usepackage{graphicx}
\usepackage{booktabs}
\usepackage{amssymb,bm,mathrsfs,bbm,amscd}
\usepackage[tbtags]{amsmath}
\usepackage{lastpage}
\usepackage{color}
\usepackage[dvipdfmx, bookmarks=true, colorlinks,%
            citecolor=blue, linkcolor=blue, anchorcolor=blue,
            filecolor=blue,urlcolor=blue,hypertex, %
            breaklinks=true]{hyperref}

\begin{document}
\begin{CJK*}{GBK}{song}



\title{Improved phenomenological nuclear charge radius formulae with kernel ridge regression
       \thanks{Supported by National Natural Science
               Foundation of China (11875027, 11775112, 11775026, 11775099, 11975096),
               Fundamental Research Funds for the Central Universities (2021MS046)}}

\author{Jian-Qin Ma$^{1}$, \quad
      Zhen-Hua Zhang$^{1}$\email{zhzhang@ncepu.edu.cn}}%

\maketitle

\address{$^{1}$Mathematics and Physics Department,
               North China Electric Power University, Beijing 102206, China}

\end{CJK*}

\begin{abstract}
The kernel ridge regression (KRR) method with Gaussian kernel is used to improve the
description of the nuclear charge radius by several phenomenological formulae.
The widely used $A^{1/3}$, $N^{1/3}$ and $Z^{1/3}$ formulae, and their
improved versions by considering the isospin dependence are adopted as examples.
The parameters in these six formulae are refitted using the Levenberg-Marquardt method,
which give better results than the previous ones.
The radius for each nucleus is predicted with the KRR network,
which is trained with the deviations between experimental and calculated
nuclear charge radii.
For each formula, the resultant root-mean-square deviations of 884 nuclei
with proton number $Z \geq 8$ and neutron number $N \geq 8$ can be reduced
to about 0.017~fm after considering the modification of the KRR method.
The extrapolation ability of the KRR method for the neutron-rich region is
examined carefully and compared with the radial basis function method.
It is found that the improved nuclear charge radius formulae by KRR method
can avoid the risk of overfitting and have a good extrapolation ability.
The influence of the ridge penalty term on the extrapolation ability
of the KRR method is also discussed.
At last, the nuclear charge radii of several recently observed K and Ca
isotopes have been analyzed.
\end{abstract}




\maketitle

\begin{multicols}{2}

\section{Introduction\label{Sec:intro}}

The nuclear charge radius, which can reflect the nuclear charge density
distribution and the Coulomb potential,
is one of the most fundamental properties of the atomic nuclei.
It depends sensitively on the properties of nuclear force and
plays a key role in investigating nuclear structure
such as shape coexistence and shape transition~\cite{Wood1992_PR215-101, Cejnar2010_RMP82-2155},
shell evolution~\cite{Thibault1981_PRC23-2720, Fricke1995_ADNDT60-177, Gorges2019_PRL122-192502},
and the nuclear volume properties connected
with exotic phenomena such as skin and halo~\cite{Tanihata1985_PRL55-2676,
Tanihata2013_PPNP68-215, Meng2015_JPG42-093101}, etc.
The accurately nuclear charge radii are also needed in many theoretical studies,
such as understanding the origin of elements
in the universe~\cite{Burbidge1957_RMP29-547, Cowan2021_RMP93-015002}.

Experimentally, considerable efforts have been devoted to the
measurement of nuclear charge radii.
By using several techniques~\cite{Cheal2010_JPG37-113101, Campbell2016_PPNP86-127},
e.g., muonic atom $x$-ray spectra, electron elastic scattering experiments, and isotope shifts,
more than 900 nuclear charge radii are provided by experiments~\cite{Angeli2013_ADNDT99-69}.
Very recently, the observation of the charge radii of several very exotic nuclei has aroused people's
attention~\cite{Ruiz2016_NatPhys12-594, Miller2019_NatPhys15-432,
Groote2020_NatPhys16-620, Koszorus2021_NatPhys17-439},
which provides a stringent test for various nuclear models.

From theoretical aspects, various methods have been developed to calculate the nuclear charge radii, e.g.,
phenomenological formulae~\cite{Bohr1969_Book, Zeng1957_ActaPhysSin13-357,
Nerlo-Pomorska1993_ZPA344-359, Duflo1994_NPA576-29, Zhang2002_EPJA13-285,
Lei2009_CTP51-123, Wang2013_PRC88-011301, Bayram2013_APPB44-1791},
macroscopic-microscopic models~\cite{Buchinger1994_PRC49-1402, Buchinger2001_PRC64-067303,
Buchinger2005_PRC72-057305, Iimura2008_PRC78-067301},
relativistic~\cite{Lalazissis1999_ADNDT71-1, Geng2005_PTP113-785, Zhao2010_PRC82-054319,
Xia2018_ADNDT121-122-1, Zhang2020_PRC102-024314, An2020_PRC102-024307,
Perera2021_PRC104-064313, Zhang2022_ADNDT144-101488}
and non-relativistic~\cite{Stoitsov2003_PRC68-054312, Goriely2009_PRL102-242501,
Goriely2010_PRC82-035804} mean-field models,
local-relation-based models~\cite{Piekarewicz2010_EPJA46-379, Sun2014_PRC90-054318,
Bao2016_PRC94-064315, Sun2017_PRC95-014307, Bao2020_PRC102-014306, Ma2021_PRC104-014303},
and $ab$ initio no-core shell model~\cite{Forssen2009_PRC79-021303}.
All of these models can provide global quantitative descriptions for
the nuclear charge radii in a wide region of nuclear chart.
However, except those local-relation-based models, the root-mean-square (rms) deviations are
larger than 0.02~fm for all of these methods, which need further improvement.

In recent years, machine learning (ML) has been employed to further improve
the accuracies of nuclear models due to its powerful and convenient inference abilities.
Various ML approaches have been adopted
to improve the description of the nuclear charge radii, e.g.,
the feed-forward neural network~\cite{Akkoyun2013_JPG40-055106, Wu2020_PRC102-054323},
the Bayesian neural network approach~\cite{Utama2016_JPG43-114002,
Neufcourt2018_PRC98-034318, Ma2020_PRC101-014304, Dong2022_PRC105-014308}, etc.
By training the ML network with the deviations between experimental and calculated
charge radii, ML approaches can reduce the corresponding
rms deviations significantly to about 0.02~fm.

In this paper, the kernel ridge regression (KRR) method with Gaussian kernel,
which is one of the most popular ML approaches, is used to improve the description
of the nuclear charge radius by taking six phenomenological formulae as examples.
Least-square fitting based on the Levenberg-Marquardt (LM) method~\cite{Marquardt1963_JSIAM11-431}
is applied in order to obtain the new parameters in these formulae,
and then the KRR method is adopted to train the charge radius residuals.
The two hyperparameters ($\sigma, \lambda$) in the KRR method
are determined by the leave-one-out cross-validation.
The performance and reliability of the extrapolated predictions
of KRR method are also analyzed in detail.
The comparison with the radial basis function (RBF) method has also been discussed,
which has been widely used to predict the nuclear
mass and $\beta$-decay half-lives, etc~\cite{Wang2011_PRC84-051303R, Niu2013_PRC88-024325,
Zheng2014_PRC90-014303, Niu2016_PRC94-054315, Niu2018_SciBull63-759,
Niu2018_PLB778-48, Shi2021_ChinPhysC45-044103}.
Note that the KRR method has already provided successful descriptions for
the nuclear mass predictions~\cite{Wu2020_PRC101-051301, Wu2021_PLB819-136387}
and also has been used to build the nuclear energy density functionals~\cite{wu2021_arxiv:2105.07696}.

This paper is organized as follows.
A brief introduction of the phenomenological nuclear charge radius formulae
and the KRR method is presented in Sec.~2.
The results obtained by the KRR method and the extrapolation power comparison to
the RBF method are given in Sec.~3.
A summary of this work is given in Sec.~4.

\section{Theoretical framework\label{Sec:theor}}

Considering the nuclear saturation property,
the radius of nuclear charge distribution is usually described by the
$A^{1/3}$ law~\cite{Bohr1969_Book}
\begin{eqnarray}
R_c=r_A A^{1/3} \ , \label{eq:A1}
\end{eqnarray}
where $A$ is the mass number and $R_c=\sqrt{5/3}\langle r^2 \rangle^{1/2}$,
with $\langle r^2 \rangle^{1/2}$ the rms nuclear charge radius.
In order to obtain the global description of the charge radius,
the parameter $r_A$ is fitted to the experimental data~\cite{Angeli2013_ADNDT99-69}.
However, it is found that the $A^{1/3}$ formula is not valid for all nuclei
since $r_A$ is not a constant but decreases systematically with increasing mass number.
Investigations show that $r_A \approx 1.30$~fm for light nuclei
and 1.20~fm for heavy nuclei.
In Ref.~\cite{Zeng1957_ActaPhysSin13-357}, a $Z^{1/3}$ law was proposed
\begin{eqnarray}
R_c=r_Z Z^{1/3} \ ,  \label{eq:Z1}
\end{eqnarray}
which is much better than the conventional $A^{1/3}$ formula.
Investigations show that the parameter $r_Z$ remains almost
a constant, i.e., $r_Z \approx 1.65$~fm, for the nuclei with $A \geq 40$.
Furthermore, the $N$-dependence of nuclear charge radii was
discussed~\cite{Angeli2009_JPG36-085102} and an $N^{1/3}$ formula
was proposed~\cite{Bayram2013_APPB44-1791}, which can be written as
\begin{eqnarray}
R_c=r_N N^{1/3} \ .  \label{eq:N1}
\end{eqnarray}

To have a better description of the nuclear charge radii,
the improved $A^{1/3}$, $N^{1/3}$ and $Z^{1/3}$ formulae
by considering the isospin dependence~\cite{Nerlo-Pomorska1993_ZPA344-359,
Bayram2013_APPB44-1791, Zhang2002_EPJA13-285}
have also been proposed, which can be written as
\begin{eqnarray}
R_c&=&r_A \left(1-b\frac{N-Z}{A}\right)A^{1/3} \ , \\
R_c&=&r_N \left(1-b\frac{N-Z}{N}\right)N^{1/3} \ , \\
R_c&=&r_Z \left(1+\frac{5}{8\pi}\beta^2\right)\left(1+b\frac{N-N^\ast}{Z}\right)Z^{1/3} \ , \label{eq:Z2}
\end{eqnarray}
where $\beta$ is the quadrupole deformation,
which is taken from~\cite{Moller2016_ADNDT109-110-1} in the present work,
and $N^\ast$ is the neutron number for the nuclei along the $\beta$-stability line,
which can be extracted from the nuclear mass
formula~\cite{Bohr1969_Book} and can be written as $Z=A/(1.98+0.0155A^{2/3})$.
$r_A$, $r_N$, $r_Z$ and $b$ are constants, which are obtained by
fitting the experimental data.
Note that the influence of deformation on nuclear charge radius
was studied systematically in Ref.~\cite{Li2021_ADNDT140-101440}.

KRR is a popular ML method with the extension of ridge regression
on the nonlinearity~\cite{Kim2012_IEEE42-1011, Wu2017_IEEE47-3916}.
It uses kernel machine to map data into higher dimensional
space and then uses regression method to treat the data.
The KRR function $S(\bm{x}_j)$ can be written as
\begin{equation}
S(\bm{x}_j)=\sum_{i=1}^m K(\bm{x}_j, \bm{x}_i)\omega_i,
\label{eq:7}
\end{equation}
where $m$ is the number of training data,
$\bm{x}_i$ denotes the location of training data, $\omega_{i}$
are weights to be determined,
and $K(\bm{x}_j, \bm{x}_i)$ is the kernel function,
which characterizes the similarity between the data.
There are several kinds of kernels can be used in the KRR method, e.g.,
linear kernel, polynomial kernel, Gaussian kernel, etc.
In the present work, the Gaussian kernel is adopted,
\begin{equation}
K(\bm{x}_j,\bm{x}_i)=\exp\left(-\frac{||\bm{x}_i-\bm{x}_j||^2}{2\sigma^2}\right) \ ,
\label{eq:8}
\end{equation}
where $\sigma$ ($\sigma>0$) is a hyperparameter
defining the range that the kernel affects.
By minimizing the following loss function
\begin{equation}
L(\bm{\omega})=\sum_{i=1}^m [S(\bm{x}_i)-y(\bm{x}_i)]^2+\lambda||\bm{\omega}||^2 \ ,
\label{eq:9}
\end{equation}
the weights $\omega_{i}$ can be determined, where $\bm{\omega} = (\omega_{1},..., \omega_{m})$.
The hyperparameter $\lambda$ ($\lambda\geq0$) determines the regularization strength
and is adopted to reduce the risk of overfitting.
Minimizing Eq.~(\ref{eq:9}) leads to
\begin{equation}
\bm{\omega}=(\bm{K}+\lambda \bm{I})^{-1}\bm{y} \ ,
\end{equation}
where $\bm{I}$ is the identity matrix and $\bm{K}$ is the kernel matrix
with elements $K_{ij}=K(\bm{x}_j, \bm{x}_i)$.

In the present work, the KRR method is applied for nuclear charge radius predictions.
Therefore, the coordinate $\bm{x}_i$ of each nucleus is naturally chosen as
$\bm{x}_i=(N_i,Z_i)$.
The Euclidean norm
\begin{equation}
 r=||\bm{x}_i-\bm{x}_j||=\sqrt{(Z_i-Z_j)^2+(N_i-N_j)^2}
\end{equation}
is defined to be the distance between two nuclei.

\section{Results and discussion\label{Sec:results}}

\end{multicols}
\begin{center}
\tabcaption{ \label{t:1}
The parameters and the root-mean-square deviations ($\Delta_{\rm rms}$) for the six
phenomenological nuclear charge radius formulae.
The experimental data are taken from~\cite{Angeli2013_ADNDT99-69},
with proton number $Z \geq 8$ and neutron number $N \geq 8$.}
\footnotesize
\begin{tabular*}{1.0\textwidth}{c@{\extracolsep{\fill}}lclc}
\toprule
Formula & Parameters & $\Delta_{\rm rms}$ (fm) & New parameters & $\Delta_{\rm rms}$ (fm) \\
\hline
$R_c=r_A A^{1/3}$ & $r_A$=1.223~fm~\cite{Bohr1969_Book}              & 0.094 & $r_A$=1.227~fm & 0.093\\
$R_c=r_N N^{1/3}$ & $r_N$=1.472~fm~\cite{Bayram2013_APPB44-1791}     & 0.151 & $r_N$=1.470~fm & 0.151\\
$R_c=r_Z Z^{1/3}$ & $r_Z$=1.631~fm~\cite{Zeng1957_ActaPhysSin13-357} & 0.076 & $r_Z$=1.639~fm & 0.072\\
$R_c=r_A \left[1-b(N-Z)/A\right]A^{1/3}$ & $r_A$=1.269~fm;
$b=0.252$~\cite{Nerlo-Pomorska1993_ZPA344-359} & 0.068 & $r_A$=1.282~fm; $b=0.342$ & 0.065\\
$R_c=r_N \left[1-b(N-Z)/N\right]N^{1/3}$ & $r_N$=1.629~fm;
$b=0.451$~\cite{Bayram2013_APPB44-1791}        & 0.063 & $r_N$=1.623~fm; $b=0.438$ & 0.063\\
$R_c=r_Z (1+5\beta^2/8\pi)\left[1+b(N-N^\ast)/Z\right]Z^{1/3}$ & $r_Z$=1.631~fm;
$b=0.062$~\cite{Zhang2002_EPJA13-285}          & 0.057 & $r_Z$=1.634~fm; $b=0.220$ & 0.049\\
\bottomrule
\end{tabular*}
\end{center}
\begin{multicols}{2}

In this work, 884 experimental data~\cite{Angeli2013_ADNDT99-69}
with proton number $Z \geq 8$ and neutron number $N \geq 8$ have been adopted for
the least-square fitting with Levenberg-Marquardt method to obtain new parameters
in these six phenomenological nuclear charge radius formulae.
The obtained parameters and the corresponding rms deviations are shown in Table.~\ref{t:1}.
In addition, old parameters and the corresponding rms deviations
obtained by previous investigations have also been shown for comparison.
These old parameters are fitted with the experimental data in
Ref.~\cite{Angeli2004_ADNDT87-185} except those two $N^{1/3}$ formulae.
It can be seen that the rms deviations are reduced a little
when these new parameters are adopted.
By considering the isospin dependence, the descriptions are
improved a lot, especially for the $N^{1/3}$ formula.
It seems worth noting that the $Z^{1/3}$ formula with only one parameter
achieves the accuracy of charge radii comparable
to those of two-parameter $A^{1/3}$ and $N^{1/3}$ formulae with isospin dependence.
Among these six phenomenological formulae,
the $Z^{1/3}$ formula with isospin dependence can reproduce the data best with
a rms deviation of 0.049~fm.
Note that the parameters have also also be refitted in Ref.~\cite{Bayram2013_APPB44-1791},
which are quite similar with the present work except those in Eq.~(\ref{eq:Z2}).
In addition, the parameters are fitted by the rms charge radius $\langle r^2 \rangle^{1/2}$
in Ref.~\cite{Bayram2013_APPB44-1791}, which has a factor of $\sqrt{5/3}$ different
from the parameters in Table~\ref{t:1}, including these two $N^{1/3}$ formulae.

\end{multicols}
\begin{center}
\centering
\includegraphics[width=0.7\textwidth]{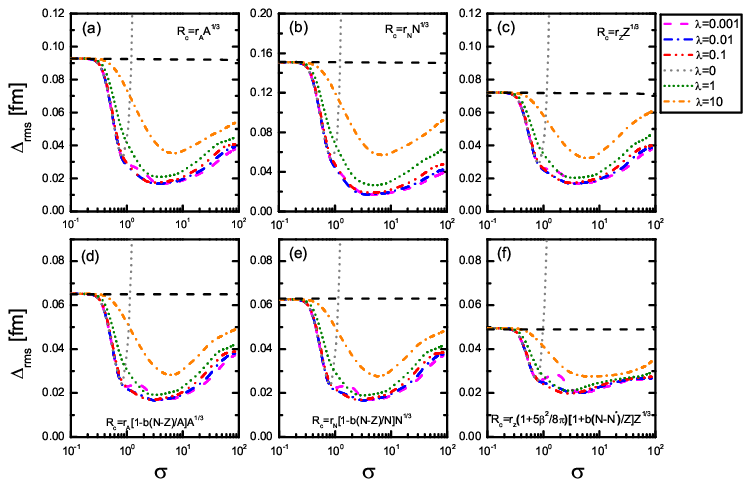}
\figcaption{\label{fig:1}
The rms deviations as a function of the
hyperparameter $\sigma$ with several selected $\lambda$ values.
For comparison, the corresponding rms deviation of each
formula is also shown with horizontal black dashed lines.}
\end{center}
\begin{multicols}{2}

The KRR function~(\ref{eq:7}) is trained to reconstruct the differences between experimental and
calculated nuclear charge radius $\Delta R(N, Z)=R^{\rm exp}(N,Z)-R^{\rm cal}(N, Z)$.
Once the weights $w_i$ are obtained, the reconstructed function
$S(N, Z)$ can be obtained for every nucleus.
Therefore, the predicted charge radius for a nucleus with neutron number $N$
and proton number $Z$ is given by $R^{\rm KRR} = R^{\rm cal}(N, Z)+S(N,Z)$.

In the present work, the leave-one-out cross-validation is adopted
to determine the hyperparameters ($\sigma, \lambda$).
In Fig.~\ref{fig:1}, the leave-one-out cross-validation rms deviations
are presented as a function of the hyperparameter $\sigma$ with
selected penalties $\lambda$ ranging from $10^{-3}$ to $10^1$.
The calculations with $\lambda=0$ are also shown,
which corresponds to the RBF results.
For comparison, the corresponding rms deviation of each
formula is also shown with horizontal black dashed lines.
It can be seen that for small $\sigma$ values, the rms deviations obtained by the KRR method
are close to those obtained by the phenomenological formulae,
regardless of the magnitudes of the $\lambda$, so the corresponding
reconstructed functions $S(N, Z)$ are quite small.
The role of the penalty term can be seen clearly with $\sigma$ increasing.
The penalty $\lambda$ has a great influence on the selection of the hyperparameter $\sigma$.
When the penalty term is neglected ($\lambda=0$), the rms deviations are minimized
at $\sigma=0.83, 0.84, 0.84, 0.82, 0.81, 0.79$ for Eqs.~(\ref{eq:A1}) to (\ref{eq:Z2}), respectively,
and they grow rapidly with increasing $\sigma$.
In addition, the rms deviations obtained with $\lambda=0$
are systematically larger than those with the penalty term $\lambda\neq0$.
It can be seen that when $\lambda\neq0$, the rms deviations
do not grow very fast for a larger $\sigma$,
in contrary to the case $\lambda=0$, which demonstrates clearly
that the penalty term can effectively prevent the results from overfitting.

\end{multicols}
\begin{center}
\tabcaption{ \label{t:2}
The adopted hyperparameters $\sigma$ and $\lambda$ by the KRR method
in each formula obtained through the leave-one-out cross-validation.
The corresponding rms deviations between the experimental data
and the KRR method are shown as $\Delta_{\rm rms}^{\rm KRR}$.
The $\Delta_{\rm rms}^{\rm CR04}$ and $\Delta_{\rm rms}^{\rm CR13-04}$ denote
the rms deviations of the training and test sets when the nuclear charge radius
in Ref.~\cite{Angeli2004_ADNDT87-185} is chosen as the training set (denoted as ``CR04''),
and the ``new'' nuclei appearing in Ref.~\cite{Angeli2013_ADNDT99-69}
is chosen as the test set (denoted as ``CR13-04'').}
\footnotesize
\begin{tabular*}{1.0\textwidth}{c@{\extracolsep{\fill}}cccccc}
\hline
\hline
Formula & $\sigma$ & $\lambda$ & $\Delta_{\rm rms}^{\rm KRR}$~(fm)  & $\Delta_{\rm rms}^{\rm CR04}$~(fm) &
$\Delta_{\rm rms}^{\rm CR13-04}$~(fm)\\
\hline
$R_c=r_A A^{1/3}$                                           & 3.01 & 0.01  & 0.0166 & 0.0125 & 0.0288   \\
$R_c=r_N N^{1/3}$                                           & 3.86 & 0.001 & 0.0165 & 0.0128 & 0.0369  \\
$R_c=r_Z Z^{1/3}$                                           & 2.93 & 0.01  & 0.0168 & 0.0123 & 0.0268  \\
$R_c=r_A \left[1-b(N-Z)/A\right]A^{1/3}$                    & 2.88 & 0.01  & 0.0165 & 0.0122 & 0.0280  \\
$R_c=r_N \left[1-b(N-Z)/N\right]N^{1/3}$                    & 2.88 & 0.01  & 0.0166 & 0.0122 & 0.0280  \\
$R_c=r_Z (1+5\beta/8\pi^2)\left[1+b(N-N^\ast)/Z\right]Z^{1/3}$ &2.46 &0.06 & 0.0197 & 0.0146 & 0.0301  \\
\bottomrule
\end{tabular*}
\end{center}
\begin{multicols}{2}

It can be seen in Fig.~\ref{fig:1} that those minima at
$\lambda=$0.001, 0.01, and 0.1 are quite close to each other.
This indicates that the results may not that sensitive to the hyperparameters in this region.
The optimized hyperparameters $\sigma$ and $\lambda$ by the KRR method
in each formula are shown in Table~\ref{t:2}.
The obtained rms deviations are smaller than 0.017~fm,
except the $Z^{1/3}$ formula with isospin dependence,
which has a rms deviation with 0.0197~fm.
It is quite interesting that this formula
can reproduce the data best without the KRR method.
However, after the KRR modification, the results are worst in these six formulae.
This may be caused by the deformation effect, since the deformation is considered only in this formula.
Maybe a better deformation parameter set can further improve the result.
It can be seen that the KRR method can enormously improve the description of the
nuclear charge radius by these phenomenological formulae,
even the original rms deviation is as large as 0.151~fm in the $N^{1/3}$ formula.
In addition, the predictive power of the KRR method was tested
by separating the nuclear charge data into two subsets,
i.e., the 782 nuclei in the nuclear charge table of 2004 (denoted as CR04)~\cite{Angeli2004_ADNDT87-185},
and the 102 ``new'' nuclei (denoted as CR13-04) appearing in Ref.~\cite{Angeli2013_ADNDT99-69}.
It can be seen that nearly all the rms deviations for the test sets are smaller than 0.03~fm,
except the $N^{1/3}$ formula.

When the hyperparameters ($\sigma, \lambda$) of each formula are determined,
the reconstructed function $S(N,Z)$ for every nuclear can be calculated by KRR method,
which are shown in the middle panels of Fig.~\ref{fig:2}.
For comparison, the differences $\Delta R=R^{\rm exp}-R^{\rm cal}$ between experimental and
the calculated values by $A^{1/3}$, $N^{1/3}$ and $Z^{1/3}$ formulae are shown in the upper panels
of Fig.~\ref{fig:2}.
The magic numbers are shown by vertical and horizontal dotted lines.
In the present work, the number of possible existing nuclei with $Z \geq 8$ and $N \geq 8$
is taken as 7275 according to Ref.~\cite{Moeller1995_ADNDT59-185}.
It can be seen clearly that for each formula, the reconstructed function $S(N,Z)$
has a similar pattern to $\Delta R$,
which indicates that the charge radius residuals can be learned well by the $S(N,Z)$.
After considering the KRR corrections, the predicted charge radius by these three formulae are in
good agreement with the data, which are shown at the lower pannels of Fig.~\ref{fig:2}.
The corresponding rms deviations are reduced to less than 0.017~fm (see Table~\ref{t:2}).

Figure~\ref{fig:3} is the same as Fig.~\ref{fig:2}, but for the three formulae
considering the isospin dependence.
It can be seen that, after considering the isospin dependence,
the descriptions of the experimental data are improved a lot,
especially for the $N^{1/3}$ formula [see Fig.~\ref{fig:3}(b)].
After considering the corrections of the reconstructed function,
the predicted charge radius (the lower panels of Fig.~\ref{fig:3})
are quite similar to those corresponding results in Fig.~\ref{fig:2}.
It can be seen in the middle panels of Figs.~\ref{fig:2} and \ref{fig:3} that
except for those nuclei close to the nuclei with known charge radius,
the KRR reconstructed function $S(N, Z)$ becomes to zero for most nuclei with unknown charge radius.
This is due to the Gaussian kernel adopted in the present calculation.
It means that for a given nucleus, very little information can be learned from the nuclei far away from it.
Therefore, for the very neutron-rich nuclei,
the reconstructed function $S(N, Z)$ vanishes
since no data can be learned from the neighboring nuclei.

\end{multicols}
\begin{center}
\centering
\includegraphics[width=0.7\textwidth]{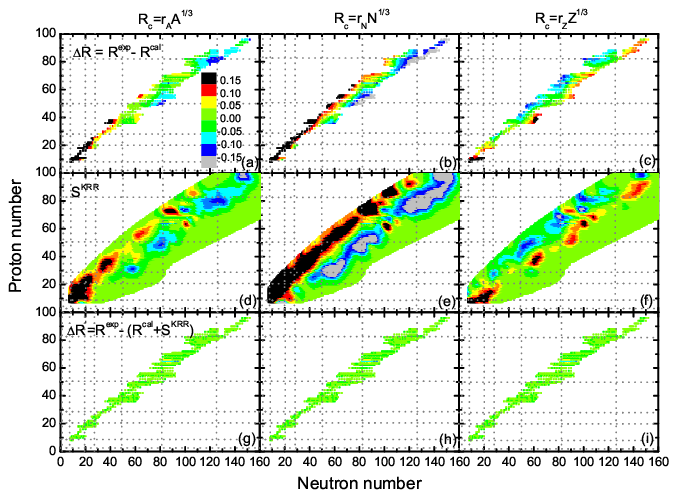}
\figcaption{\label{fig:2}
The differences $\Delta R=R^{\rm exp}-R^{\rm cal}$ between experimental and
the calculated values by $A^{1/3}$, $N^{1/3}$ and $Z^{1/3}$ formulae (upper panels),
the KRR reconstructed function $S(N, Z)$ (middle panels),
and the differences $\Delta R'=R^{\rm exp}-(R^{\rm cal} + S^{\rm KRR})$ between experimental
and the predictions of these three formulae with the KRR corrections  (lower pannels).
The magic numbers are shown by vertical and horizontal dotted lines.
The possible existing nuclei are taken from~\cite{Moeller1995_ADNDT59-185}.}
\end{center}
\begin{multicols}{2}

\end{multicols}
\vspace{-6mm}
\begin{center}
\centering
\includegraphics[width=0.7\textwidth]{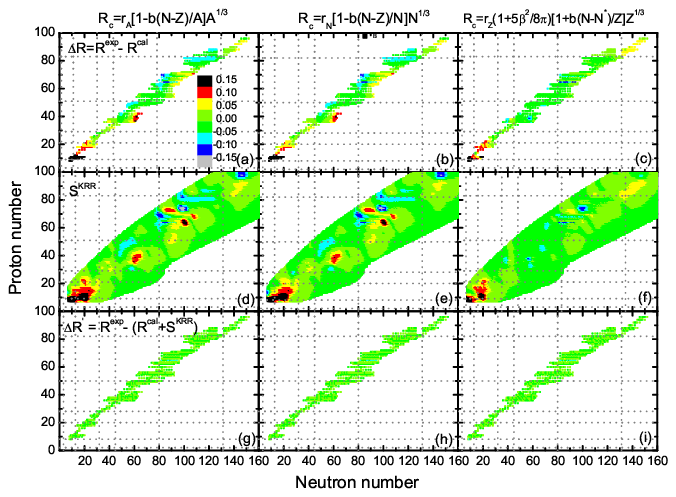}
\figcaption{\label{fig:3}The same as Fig.~\ref{fig:2}, but for the three formulae
considering the isospin dependence.
}
\end{center}
\begin{multicols}{2}

\vspace{4mm}
\begin{center}
\centering
\includegraphics[width=1.0\columnwidth]{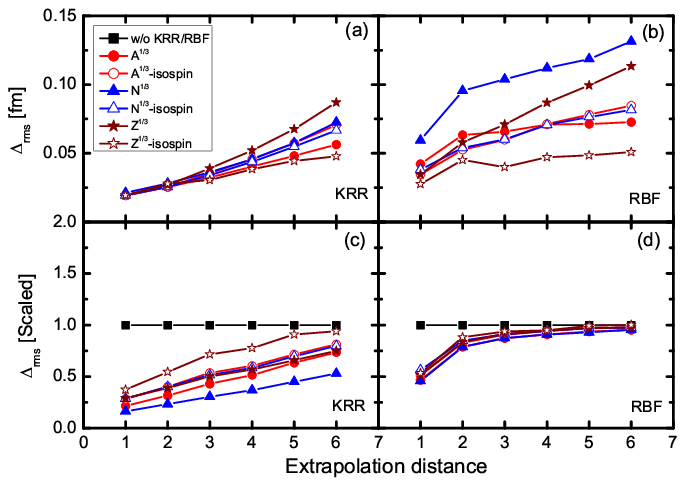}
\figcaption{\label{fig:4}
Comparison of the extrapolation power of the KRR and the RBF methods
for six test sets with different extrapolation distances.
The upper panels show the extrapolated rms deviations of
KRR and RBF method.
The lower panels show the rms deviations scaled
to the corresponding rms deviations for the phenomenological
charge radius formulae without KRR or RBF corrections.}
\end{center}

To study the predictive power of the KRR method to the neutron-rich nuclei,
the 884 nuclei with known charge radius are redivided into training set and test sets as follows.
For each isotopic chain with more than nine nuclei, the six most neutron-rich ones are removed out
from the training set, and then, they are classified into
six test sets according to the distance to the last nucleus in the training set,
i.e., the test set 1 (6) has the shortest (longest) extrapolation distance.
For comparison, predictive power of the RBF method
with the Gaussian kernel is also studied.
Note that the hyperparameters obtained by the leave-one-out cross-validation
remain the same in the following studies of KRR and RBF extrapolations.

In Figs.~\ref{fig:4}(a) and (b), the rms deviations of the calculated nuclear charge radius
after taking into account the KRR and RBF corrections are shown
as a function of the extrapolation distance for six test sets.
It can be seen that the rms deviations of these six formulae are increasing
with extrapolation distance for both KRR and RBF methods.
In the KRR method [see Fig.~\ref{fig:4}(a)], the extrapolation power of $Z^{1/3}$ formula is the worst,
while after considering the isospoin dependence, it becomes the best.
In the RBF method [see Fig.~\ref{fig:4}(b)], the extrapolation power of $N^{1/3}$ formula is the worst.
After considering the isospoin dependence, the results become much better.
The $Z^{1/3}$ formula with isospin dependence is also the best one among these six formulae.
Note that the extrapolation power of $A^{1/3}$ formula with isospin dependence
becomes worse than the traditional $A^{1/3}$ formula with larger extrapolation distance
both in KRR and RBF methods.
To see it more clearly, in Figs.~\ref{fig:4}(c) and (d), the rms deviations are scaled
to the corresponding rms deviations of the
phenomenological charge radius formulae without KRR or RBF corrections.
It can be seen in Fig.~\ref{fig:4}(c) that
the scaled rms deviations increase approximately linearly
and there is no overfitting in the KRR method for all of these six formulae.
As for the RBF method, the scaled rms deviations increase sharply from
the first to the second extrapolation step,
and the overfitting appears in the fifth or the sixth step of extrapolation.
Note that the overfitting is not that serious due to the Gaussian kernel
adopted in the present RBF calculation.
If the linear kernel is adopted, the overfitting will be much more obvious,
which has been shown in the mass prediction in Ref.~\cite{Wu2020_PRC101-051301}.
Thus, it can be seen clearly that compared with RBF method, the KRR method
has a better extrapolation power for any phenomenological formula.
This is because the ridge penalty term $\lambda$
in the KRR method can automatically identify the limit of the extrapolation distance.

\vspace{2mm}
\begin{center}
\centering
\includegraphics[width=1.0\columnwidth]{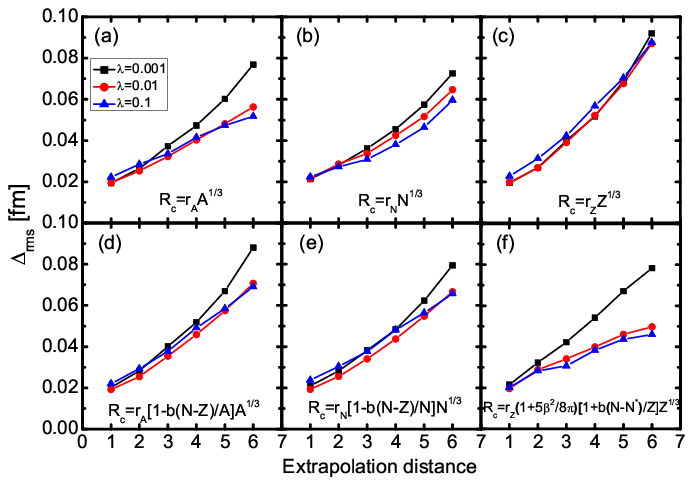}
\figcaption{\label{fig:5}
The effect of the ridge penalty term
$\lambda$ on the extrapolation power.}
\end{center}

As shown in Fig.~\ref{fig:1} before, the rms deviations are
not sensitive to the hyperparameter $\lambda$
when $\lambda$ is chosen as 0.1, 0.01 and 0.001, respectively.
Therefore, it is important to know the influence of the ridge penalty term
$\lambda$ on the extrapolation power.
Fig.~\ref{fig:5} shows the extrapolated rms deviation of KRR method
with $\lambda=$0.1, 0.01 and 0.001, respectively.
Note that the hyperparameter $\sigma$ is chosen as the optimum value
for each $\lambda$, which can obtain the smallest rms deviation.
It can be seen that for most cases, a smaller $\lambda$ gives an obviously
worse extrapolation power, especially for $\lambda=0.001$,
except the $Z^{1/3}$ formula [Fig.~\ref{fig:5}(c)].
For $\lambda=0.01$ and 0.1, the extrapolation power is quite
similar for these six formulae.
Therefore, it should be very careful to choose the
hyperparameters if they are not quite sensitive to the results.
Note that when $\lambda=$0.001 is adopted in the $N^{1/3}$ formula,
the extrapolation power is not that bad.
Therefore, the hyperparameters adopted in the present
work (see Table~\ref{t:2}) are quite reasonable.

\vspace{2mm}
\begin{center}
\centering
\includegraphics[width=1.0\columnwidth]{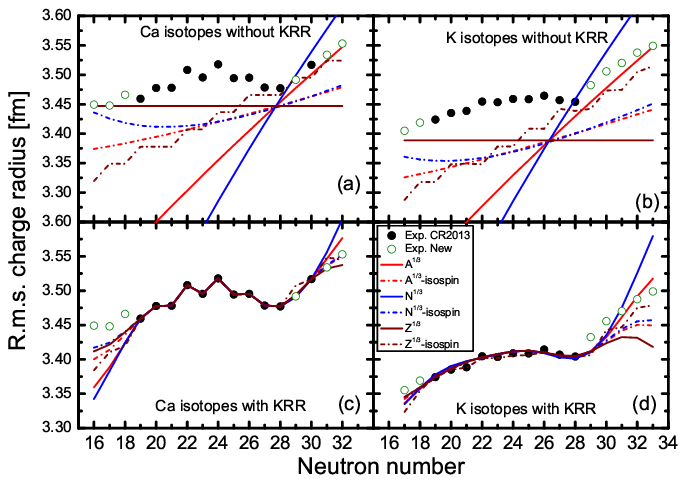}
\figcaption{\label{fig:6}
Comparison between the experimental and calculated
root-mean-square nuclear charge radii without (upper panels)
and with (lower panels) KRR corrections for Ca and K isotopes.
The experimental data taken from Ref.~\cite{Angeli2013_ADNDT99-69}
(denoted as ``CR2013'') are shown by black solid circles, and the new
data taken from Refs.~\cite{Ruiz2016_NatPhys12-594, Miller2019_NatPhys15-432,
Koszorus2021_NatPhys17-439} (denoted as ``New'') are shown by olive open circles.}
\end{center}

Very recently, the charge radii of several very exotic K and Ca isotopes
have been observed~\cite{Ruiz2016_NatPhys12-594, Miller2019_NatPhys15-432,
Koszorus2021_NatPhys17-439}.
Figure~\ref{fig:6} shows the comparison between the experimental and calculated
root-mean-square nuclear charge radii without (upper panels)
and with (lower panels) KRR corrections for Ca and K isotopes.
It can be seen that the calculated results by these six formulae
without KRR corrections deviate a lot from the experimental data.
After the KRR corrections being considered,
all these six formulae can reproduce the data in ``CR2013'' quite well.
However, for the ``new'' data observed by later experiments, the calculations become quite different.
For the Ca isotopes [Fig.~\ref{fig:6}(c)], the $N^{1/3}$ and $A^{1/3}$ formulae
reproduce the data not very well, while other formulae reproduce the data at the same level.
For the K isotopes [Fig.~\ref{fig:6}(d)], all the formulae can reproduce the
data for the proton-rich side, while the $A^{1/3}$ formula and the $Z^{1/3}$ formula
with isospin dependence can reproduce the data better for the neutron-rich side.
It is also interesting to see that only the $Z^{1/3}$ formulae
with isospin dependence can reproduce the slightly staggering in the K isotopes.
Maybe it is due to the deformation effect considered in this formula.
Therefore, it can be seen that although KRR method is a powerful machine learning
method, a microscopic model which can provide a better description of
the nuclear charge radius is still needed.
Note that the Bayesian neural network has also been applied to
study the nuclear charge radii for the Ca and K isotopes recently~\cite{Dong2022_PRC105-014308}.

\section{Summary\label{Sec:summary}}

In summary, the kernel ridge regression (KRR) method is adopted to improve the
description of the nuclear charge radius by several phenomenological formulae.
The widely used $A^{1/3}$, $N^{1/3}$ and $Z^{1/3}$ formulae, and their
improved versions by considering the isospin dependence are adopted as examples.
First, 884 experimental data with proton number $Z \geq 8$ and
neutron number $N \geq 8$ have been adopted for
the least-square fitting with Levenberg-Marquardt method to obtain new parameters
in these six phenomenological nuclear charge radius formulae.
The root-mean-square deviations are reduced
when these new parameters are adopted.
Then the radius for each nucleus is predicted with the KRR network, which is
trained with the deviations between experimental and calculated nuclear charge radii.
For each formula, the resultant root-mean-square deviations can be reduced
to about 0.017~fm after considering the modification of the KRR method.
The extrapolation ability of the KRR method for the neutron-rich region is
examined carefully and compared with the radial basis function method.
It is found that compared with the RBF method,
the improved nuclear charge radius formulae by KRR method
can avoid the risk of overfitting and have a good extrapolation ability.
The influence of the ridge penalty term on the extrapolation ability
of the KRR method is analyzed.
The charge radii of several recently observed K and Ca isotopes have also been analyzed.

\acknowledgments{The authors are grateful to X. H. Wu and P. W. Zhao for fruitful discussions.}

\end{multicols}

\vspace{-1mm}
\centerline{\rule{80mm}{0.1pt}}
\vspace{2mm}

\begin{multicols}{2}


\begin{thebibliography}{71}%
\makeatletter
\providecommand \@ifxundefined [1]{%
 \@ifx{#1\undefined}
}%
\providecommand \@ifnum [1]{%
 \ifnum #1\expandafter \@firstoftwo
 \else \expandafter \@secondoftwo
 \fi
}%
\providecommand \@ifx [1]{%
 \ifx #1\expandafter \@firstoftwo
 \else \expandafter \@secondoftwo
 \fi
}%
\providecommand \natexlab [1]{#1}%
\providecommand \enquote  [1]{``#1''}%
\providecommand \bibnamefont  [1]{#1}%
\providecommand \bibfnamefont [1]{#1}%
\providecommand \citenamefont [1]{#1}%
\providecommand \href@noop [0]{\@secondoftwo}%
\providecommand \href [0]{\begingroup \@sanitize@url \@href}%
\providecommand \@href[1]{\@@startlink{#1}\@@href}%
\providecommand \@@href[1]{\endgroup#1\@@endlink}%
\providecommand \@sanitize@url [0]{\catcode `\\12\catcode `\$12\catcode
  `\&12\catcode `\#12\catcode `\^12\catcode `\_12\catcode `\%12\relax}%
\providecommand \@@startlink[1]{}%
\providecommand \@@endlink[0]{}%
\providecommand \url  [0]{\begingroup\@sanitize@url \@url }%
\providecommand \@url [1]{\endgroup\@href {#1}{\urlprefix }}%
\providecommand \urlprefix  [0]{URL }%
\providecommand \Eprint [0]{\href }%
\providecommand \doibase [0]{http://dx.doi.org/}%
\providecommand \selectlanguage [0]{\@gobble}%
\providecommand \bibinfo  [0]{\@secondoftwo}%
\providecommand \bibfield  [0]{\@secondoftwo}%
\providecommand \translation [1]{[#1]}%
\providecommand \BibitemOpen [0]{}%
\providecommand \bibitemStop [0]{}%
\providecommand \bibitemNoStop [0]{.\EOS\space}%
\providecommand \EOS [0]{\spacefactor3000\relax}%
\providecommand \BibitemShut  [1]{\csname bibitem#1\endcsname}%
\let\auto@bib@innerbib\@empty
\bibitem [{\citenamefont {Wood}\ \emph {et~al.}(1992)\citenamefont {Wood},
  \citenamefont {Heyde}, \citenamefont {Nazarewicz}, \citenamefont {Huyse},\
  and\ \citenamefont {van Duppen}}]{Wood1992_PR215-101}%
  \BibitemOpen
  \bibfield  {author} {\bibinfo {author} {\bibfnamefont {J.~L.}\ \bibnamefont
  {Wood}}, \bibinfo {author} {\bibfnamefont {K.}~\bibnamefont {Heyde}},
  \bibinfo {author} {\bibfnamefont {W.}~\bibnamefont {Nazarewicz}}, \bibinfo
  {author} {\bibfnamefont {M.}~\bibnamefont {Huyse}}, \ and\ \bibinfo {author}
  {\bibfnamefont {P.}~\bibnamefont {van Duppen}},\ }\href {\doibase
  10.1016/0370-1573(92)90095-H} {\bibfield  {journal} {\bibinfo  {journal}
  {Phys. Rep.}\ }\textbf {\bibinfo {volume} {215}},\ \bibinfo {pages} {101}
  (\bibinfo {year} {1992})}\BibitemShut {NoStop}%
\bibitem [{\citenamefont {Cejnar}\ \emph {et~al.}(2010)\citenamefont {Cejnar},
  \citenamefont {Jolie},\ and\ \citenamefont {Casten}}]{Cejnar2010_RMP82-2155}%
  \BibitemOpen
  \bibfield  {author} {\bibinfo {author} {\bibfnamefont {P.}~\bibnamefont
  {Cejnar}}, \bibinfo {author} {\bibfnamefont {J.}~\bibnamefont {Jolie}}, \
  and\ \bibinfo {author} {\bibfnamefont {R.~F.}\ \bibnamefont {Casten}},\
  }\href {\doibase 10.1103/RevModPhys.82.2155} {\bibfield  {journal} {\bibinfo
  {journal} {Rev. Mod. Phys.}\ }\textbf {\bibinfo {volume} {82}},\ \bibinfo
  {pages} {2155} (\bibinfo {year} {2010})}\BibitemShut {NoStop}%
\bibitem [{\citenamefont {Thibault}\ \emph {et~al.}(1981)\citenamefont
  {Thibault}, \citenamefont {Touchard}, \citenamefont {B\"uttgenbach},
  \citenamefont {Klapisch}, \citenamefont {de~Saint~Simon}, \citenamefont
  {Duong}, \citenamefont {Jacquinot}, \citenamefont {Juncar}, \citenamefont
  {Liberman}, \citenamefont {Pillet}, \citenamefont {Pinard}, \citenamefont
  {Vialle}, \citenamefont {Pesnelle},\ and\ \citenamefont
  {Huber}}]{Thibault1981_PRC23-2720}%
  \BibitemOpen
  \bibfield  {author} {\bibinfo {author} {\bibfnamefont {C.}~\bibnamefont
  {Thibault}}, \bibinfo {author} {\bibfnamefont {F.}~\bibnamefont {Touchard}},
  \bibinfo {author} {\bibfnamefont {S.}~\bibnamefont {B\"uttgenbach}}, \bibinfo
  {author} {\bibfnamefont {R.}~\bibnamefont {Klapisch}}, \bibinfo {author}
  {\bibfnamefont {M.}~\bibnamefont {de~Saint~Simon}}, \bibinfo {author}
  {\bibfnamefont {H.~T.}\ \bibnamefont {Duong}}, \bibinfo {author}
  {\bibfnamefont {P.}~\bibnamefont {Jacquinot}}, \bibinfo {author}
  {\bibfnamefont {P.}~\bibnamefont {Juncar}}, \bibinfo {author} {\bibfnamefont
  {S.}~\bibnamefont {Liberman}}, \bibinfo {author} {\bibfnamefont
  {P.}~\bibnamefont {Pillet}}, \bibinfo {author} {\bibfnamefont
  {J.}~\bibnamefont {Pinard}}, \bibinfo {author} {\bibfnamefont {J.~L.}\
  \bibnamefont {Vialle}}, \bibinfo {author} {\bibfnamefont {A.}~\bibnamefont
  {Pesnelle}}, \ and\ \bibinfo {author} {\bibfnamefont {G.}~\bibnamefont
  {Huber}},\ }\href {\doibase 10.1103/PhysRevC.23.2720} {\bibfield  {journal}
  {\bibinfo  {journal} {Phys. Rev. C}\ }\textbf {\bibinfo {volume} {23}},\
  \bibinfo {pages} {2720} (\bibinfo {year} {1981})}\BibitemShut {NoStop}%
\bibitem [{\citenamefont {Fricke}\ \emph {et~al.}(1995)\citenamefont {Fricke},
  \citenamefont {Bernhardt}, \citenamefont {Heilig}, \citenamefont {Schaller},
  \citenamefont {Schellenberg}, \citenamefont {Shera},\ and\ \citenamefont
  {Dejager}}]{Fricke1995_ADNDT60-177}%
  \BibitemOpen
  \bibfield  {author} {\bibinfo {author} {\bibfnamefont {G.}~\bibnamefont
  {Fricke}}, \bibinfo {author} {\bibfnamefont {C.}~\bibnamefont {Bernhardt}},
  \bibinfo {author} {\bibfnamefont {K.}~\bibnamefont {Heilig}}, \bibinfo
  {author} {\bibfnamefont {L.}~\bibnamefont {Schaller}}, \bibinfo {author}
  {\bibfnamefont {L.}~\bibnamefont {Schellenberg}}, \bibinfo {author}
  {\bibfnamefont {E.}~\bibnamefont {Shera}}, \ and\ \bibinfo {author}
  {\bibfnamefont {C.}~\bibnamefont {Dejager}},\ }\href {\doibase
  10.1006/adnd.1995.1007} {\bibfield  {journal} {\bibinfo  {journal} {At. Data
  Nucl. Data Tables}\ }\textbf {\bibinfo {volume} {60}},\ \bibinfo {pages}
  {177} (\bibinfo {year} {1995})}\BibitemShut {NoStop}%
\bibitem [{\citenamefont {Gorges}\ \emph {et~al.}(2019)\citenamefont {Gorges},
  \citenamefont {Rodr\'{\i}guez}, \citenamefont {Balabanski}, \citenamefont
  {Bissell}, \citenamefont {Blaum}, \citenamefont {Cheal}, \citenamefont
  {Garcia~Ruiz}, \citenamefont {Georgiev}, \citenamefont {Gins}, \citenamefont
  {Heylen}, \citenamefont {Kanellakopoulos}, \citenamefont {Kaufmann},
  \citenamefont {Kowalska}, \citenamefont {Lagaki}, \citenamefont {Lechner},
  \citenamefont {Maa\ss{}}, \citenamefont {Malbrunot-Ettenauer}, \citenamefont
  {Nazarewicz}, \citenamefont {Neugart}, \citenamefont {Neyens}, \citenamefont
  {N\"ortersh\"auser}, \citenamefont {Reinhard}, \citenamefont {Sailer},
  \citenamefont {S\'anchez}, \citenamefont {Schmidt}, \citenamefont {Wehner},
  \citenamefont {Wraith}, \citenamefont {Xie}, \citenamefont {Xu},
  \citenamefont {Yang},\ and\ \citenamefont
  {Yordanov}}]{Gorges2019_PRL122-192502}%
  \BibitemOpen
  \bibfield  {author} {\bibinfo {author} {\bibfnamefont {C.}~\bibnamefont
  {Gorges}}, \bibinfo {author} {\bibfnamefont {L.~V.}\ \bibnamefont
  {Rodr\'{\i}guez}}, \bibinfo {author} {\bibfnamefont {D.~L.}\ \bibnamefont
  {Balabanski}}, \bibinfo {author} {\bibfnamefont {M.~L.}\ \bibnamefont
  {Bissell}}, \bibinfo {author} {\bibfnamefont {K.}~\bibnamefont {Blaum}},
  \bibinfo {author} {\bibfnamefont {B.}~\bibnamefont {Cheal}}, \bibinfo
  {author} {\bibfnamefont {R.~F.}\ \bibnamefont {Garcia~Ruiz}}, \bibinfo
  {author} {\bibfnamefont {G.}~\bibnamefont {Georgiev}}, \bibinfo {author}
  {\bibfnamefont {W.}~\bibnamefont {Gins}}, \bibinfo {author} {\bibfnamefont
  {H.}~\bibnamefont {Heylen}}, \bibinfo {author} {\bibfnamefont
  {A.}~\bibnamefont {Kanellakopoulos}}, \bibinfo {author} {\bibfnamefont
  {S.}~\bibnamefont {Kaufmann}}, \bibinfo {author} {\bibfnamefont
  {M.}~\bibnamefont {Kowalska}}, \bibinfo {author} {\bibfnamefont
  {V.}~\bibnamefont {Lagaki}}, \bibinfo {author} {\bibfnamefont
  {S.}~\bibnamefont {Lechner}}, \bibinfo {author} {\bibfnamefont
  {B.}~\bibnamefont {Maa\ss{}}}, \bibinfo {author} {\bibfnamefont
  {S.}~\bibnamefont {Malbrunot-Ettenauer}}, \bibinfo {author} {\bibfnamefont
  {W.}~\bibnamefont {Nazarewicz}}, \bibinfo {author} {\bibfnamefont
  {R.}~\bibnamefont {Neugart}}, \bibinfo {author} {\bibfnamefont
  {G.}~\bibnamefont {Neyens}}, \bibinfo {author} {\bibfnamefont
  {W.}~\bibnamefont {N\"ortersh\"auser}}, \bibinfo {author} {\bibfnamefont
  {P.-G.}\ \bibnamefont {Reinhard}}, \bibinfo {author} {\bibfnamefont
  {S.}~\bibnamefont {Sailer}}, \bibinfo {author} {\bibfnamefont
  {R.}~\bibnamefont {S\'anchez}}, \bibinfo {author} {\bibfnamefont
  {S.}~\bibnamefont {Schmidt}}, \bibinfo {author} {\bibfnamefont
  {L.}~\bibnamefont {Wehner}}, \bibinfo {author} {\bibfnamefont
  {C.}~\bibnamefont {Wraith}}, \bibinfo {author} {\bibfnamefont
  {L.}~\bibnamefont {Xie}}, \bibinfo {author} {\bibfnamefont {Z.~Y.}\
  \bibnamefont {Xu}}, \bibinfo {author} {\bibfnamefont {X.~F.}\ \bibnamefont
  {Yang}}, \ and\ \bibinfo {author} {\bibfnamefont {D.~T.}\ \bibnamefont
  {Yordanov}},\ }\href {\doibase 10.1103/PhysRevLett.122.192502} {\bibfield
  {journal} {\bibinfo  {journal} {Phys. Rev. Lett.}\ }\textbf {\bibinfo
  {volume} {122}},\ \bibinfo {pages} {192502} (\bibinfo {year}
  {2019})}\BibitemShut {NoStop}%
\bibitem [{\citenamefont {Tanihata}\ \emph {et~al.}(1985)\citenamefont
  {Tanihata}, \citenamefont {Hamagaki}, \citenamefont {Hashimoto},
  \citenamefont {Shida}, \citenamefont {Yoshikawa}, \citenamefont {Sugimoto},
  \citenamefont {Yamakawa}, \citenamefont {Kobayashi},\ and\ \citenamefont
  {Takahashi}}]{Tanihata1985_PRL55-2676}%
  \BibitemOpen
  \bibfield  {author} {\bibinfo {author} {\bibfnamefont {I.}~\bibnamefont
  {Tanihata}}, \bibinfo {author} {\bibfnamefont {H.}~\bibnamefont {Hamagaki}},
  \bibinfo {author} {\bibfnamefont {O.}~\bibnamefont {Hashimoto}}, \bibinfo
  {author} {\bibfnamefont {Y.}~\bibnamefont {Shida}}, \bibinfo {author}
  {\bibfnamefont {N.}~\bibnamefont {Yoshikawa}}, \bibinfo {author}
  {\bibfnamefont {K.}~\bibnamefont {Sugimoto}}, \bibinfo {author}
  {\bibfnamefont {O.}~\bibnamefont {Yamakawa}}, \bibinfo {author}
  {\bibfnamefont {T.}~\bibnamefont {Kobayashi}}, \ and\ \bibinfo {author}
  {\bibfnamefont {N.}~\bibnamefont {Takahashi}},\ }\href {\doibase
  10.1103/PhysRevLett.55.2676} {\bibfield  {journal} {\bibinfo  {journal}
  {Phys. Rev. Lett.}\ }\textbf {\bibinfo {volume} {55}},\ \bibinfo {pages}
  {2676} (\bibinfo {year} {1985})}\BibitemShut {NoStop}%
\bibitem [{\citenamefont {Tanihata}\ \emph {et~al.}(2013)\citenamefont
  {Tanihata}, \citenamefont {Savajols},\ and\ \citenamefont
  {Kanungo}}]{Tanihata2013_PPNP68-215}%
  \BibitemOpen
  \bibfield  {author} {\bibinfo {author} {\bibfnamefont {I.}~\bibnamefont
  {Tanihata}}, \bibinfo {author} {\bibfnamefont {H.}~\bibnamefont {Savajols}},
  \ and\ \bibinfo {author} {\bibfnamefont {R.}~\bibnamefont {Kanungo}},\ }\href
  {\doibase 10.1016/j.ppnp.2012.07.001} {\bibfield  {journal} {\bibinfo
  {journal} {Prog. Part. Nucl. Phys.}\ }\textbf {\bibinfo {volume} {68}},\
  \bibinfo {pages} {215} (\bibinfo {year} {2013})}\BibitemShut {NoStop}%
\bibitem [{\citenamefont {Meng}\ and\ \citenamefont
  {Zhou}(2015)}]{Meng2015_JPG42-093101}%
  \BibitemOpen
  \bibfield  {author} {\bibinfo {author} {\bibfnamefont {J.}~\bibnamefont
  {Meng}}\ and\ \bibinfo {author} {\bibfnamefont {S.~G.}\ \bibnamefont
  {Zhou}},\ }\href {\doibase 10.1088/0954-3899/42/9/093101} {\bibfield
  {journal} {\bibinfo  {journal} {J. Phys. G: Nucl. Part. Phys.}\ }\textbf
  {\bibinfo {volume} {42}},\ \bibinfo {pages} {093101} (\bibinfo {year}
  {2015})}\BibitemShut {NoStop}%
\bibitem [{\citenamefont {Burbidge}\ \emph {et~al.}(1957)\citenamefont
  {Burbidge}, \citenamefont {Burbidge}, \citenamefont {Fowler},\ and\
  \citenamefont {Hoyle}}]{Burbidge1957_RMP29-547}%
  \BibitemOpen
  \bibfield  {author} {\bibinfo {author} {\bibfnamefont {E.~M.}\ \bibnamefont
  {Burbidge}}, \bibinfo {author} {\bibfnamefont {G.~R.}\ \bibnamefont
  {Burbidge}}, \bibinfo {author} {\bibfnamefont {W.~A.}\ \bibnamefont
  {Fowler}}, \ and\ \bibinfo {author} {\bibfnamefont {F.}~\bibnamefont
  {Hoyle}},\ }\href {\doibase 10.1103/RevModPhys.29.547} {\bibfield  {journal}
  {\bibinfo  {journal} {Rev. Mod. Phys.}\ }\textbf {\bibinfo {volume} {29}},\
  \bibinfo {pages} {547} (\bibinfo {year} {1957})}\BibitemShut {NoStop}%
\bibitem [{\citenamefont {Cowan}\ \emph {et~al.}(2021)\citenamefont {Cowan},
  \citenamefont {Sneden}, \citenamefont {Lawler}, \citenamefont {Aprahamian},
  \citenamefont {Wiescher}, \citenamefont {Langanke}, \citenamefont
  {Mart\'{\i}nez-Pinedo},\ and\ \citenamefont
  {Thielemann}}]{Cowan2021_RMP93-015002}%
  \BibitemOpen
  \bibfield  {author} {\bibinfo {author} {\bibfnamefont {J.~J.}\ \bibnamefont
  {Cowan}}, \bibinfo {author} {\bibfnamefont {C.}~\bibnamefont {Sneden}},
  \bibinfo {author} {\bibfnamefont {J.~E.}\ \bibnamefont {Lawler}}, \bibinfo
  {author} {\bibfnamefont {A.}~\bibnamefont {Aprahamian}}, \bibinfo {author}
  {\bibfnamefont {M.}~\bibnamefont {Wiescher}}, \bibinfo {author}
  {\bibfnamefont {K.}~\bibnamefont {Langanke}}, \bibinfo {author}
  {\bibfnamefont {G.}~\bibnamefont {Mart\'{\i}nez-Pinedo}}, \ and\ \bibinfo
  {author} {\bibfnamefont {F.-K.}\ \bibnamefont {Thielemann}},\ }\href
  {\doibase 10.1103/RevModPhys.93.015002} {\bibfield  {journal} {\bibinfo
  {journal} {Rev. Mod. Phys.}\ }\textbf {\bibinfo {volume} {93}},\ \bibinfo
  {pages} {015002} (\bibinfo {year} {2021})}\BibitemShut {NoStop}%
\bibitem [{\citenamefont {Cheal}\ and\ \citenamefont
  {Flanagan}(2010)}]{Cheal2010_JPG37-113101}%
  \BibitemOpen
  \bibfield  {author} {\bibinfo {author} {\bibfnamefont {B.}~\bibnamefont
  {Cheal}}\ and\ \bibinfo {author} {\bibfnamefont {K.~T.}\ \bibnamefont
  {Flanagan}},\ }\href {\doibase 10.1088/0954-3899/37/11/113101} {\bibfield
  {journal} {\bibinfo  {journal} {J. Phys. G: Nucl. Part. Phys.}\ }\textbf
  {\bibinfo {volume} {37}},\ \bibinfo {pages} {113101} (\bibinfo {year}
  {2010})}\BibitemShut {NoStop}%
\bibitem [{\citenamefont {Campbell}\ \emph {et~al.}(2016)\citenamefont
  {Campbell}, \citenamefont {Moore},\ and\ \citenamefont
  {Pearson}}]{Campbell2016_PPNP86-127}%
  \BibitemOpen
  \bibfield  {author} {\bibinfo {author} {\bibfnamefont {P.}~\bibnamefont
  {Campbell}}, \bibinfo {author} {\bibfnamefont {I.~D.}\ \bibnamefont {Moore}},
  \ and\ \bibinfo {author} {\bibfnamefont {M.~R.}\ \bibnamefont {Pearson}},\
  }\href {\doibase 10.1016/j.ppnp.2015.09.003} {\bibfield  {journal} {\bibinfo
  {journal} {Prog. Part. Nucl. Phys.}\ }\textbf {\bibinfo {volume} {86}},\
  \bibinfo {pages} {127} (\bibinfo {year} {2016})}\BibitemShut {NoStop}%
\bibitem [{\citenamefont {Angeli}\ and\ \citenamefont
  {Marinova}(2013)}]{Angeli2013_ADNDT99-69}%
  \BibitemOpen
  \bibfield  {author} {\bibinfo {author} {\bibfnamefont {I.}~\bibnamefont
  {Angeli}}\ and\ \bibinfo {author} {\bibfnamefont {K.}~\bibnamefont
  {Marinova}},\ }\href {\doibase http://dx.doi.org/10.1016/j.adt.2011.12.006}
  {\bibfield  {journal} {\bibinfo  {journal} {At. Data Nucl. Data Tables}\
  }\textbf {\bibinfo {volume} {99}},\ \bibinfo {pages} {69} (\bibinfo {year}
  {2013})}\BibitemShut {NoStop}%
\bibitem [{\citenamefont {Ruiz}\ \emph {et~al.}(2016)\citenamefont {Ruiz},
  \citenamefont {F.}, \citenamefont {Bissell}, \citenamefont {Blaum},
  \citenamefont {Ekstr\"om}, \citenamefont {Fr\"ommgen}, \citenamefont {Hagen},
  \citenamefont {Hammen}, \citenamefont {Hebeler}, \citenamefont {Holt},
  \citenamefont {Jansen}, \citenamefont {Kowalska}, \citenamefont {Kreim},
  \citenamefont {Nazarewicz}, \citenamefont {Neugart}, \citenamefont {Neyens},
  \citenamefont {N\"ortersh\"auser}, \citenamefont {Papenbrock}, \citenamefont
  {Papuga}, \citenamefont {Schwenk}, \citenamefont {Simonis}, \citenamefont
  {Wendt},\ and\ \citenamefont {Yordanov}}]{Ruiz2016_NatPhys12-594}%
  \BibitemOpen
  \bibfield  {author} {\bibinfo {author} {\bibfnamefont {G.}~\bibnamefont
  {Ruiz}}, \bibinfo {author} {\bibfnamefont {R.}~\bibnamefont {F.}}, \bibinfo
  {author} {\bibfnamefont {M.~L.}\ \bibnamefont {Bissell}}, \bibinfo {author}
  {\bibfnamefont {K.}~\bibnamefont {Blaum}}, \bibinfo {author} {\bibfnamefont
  {A.}~\bibnamefont {Ekstr\"om}}, \bibinfo {author} {\bibfnamefont
  {N.}~\bibnamefont {Fr\"ommgen}}, \bibinfo {author} {\bibfnamefont
  {G.}~\bibnamefont {Hagen}}, \bibinfo {author} {\bibfnamefont
  {M.}~\bibnamefont {Hammen}}, \bibinfo {author} {\bibfnamefont
  {K.}~\bibnamefont {Hebeler}}, \bibinfo {author} {\bibfnamefont {J.~D.}\
  \bibnamefont {Holt}}, \bibinfo {author} {\bibfnamefont {G.~R.}\ \bibnamefont
  {Jansen}}, \bibinfo {author} {\bibfnamefont {M.}~\bibnamefont {Kowalska}},
  \bibinfo {author} {\bibfnamefont {K.}~\bibnamefont {Kreim}}, \bibinfo
  {author} {\bibfnamefont {W.}~\bibnamefont {Nazarewicz}}, \bibinfo {author}
  {\bibfnamefont {R.}~\bibnamefont {Neugart}}, \bibinfo {author} {\bibfnamefont
  {G.}~\bibnamefont {Neyens}}, \bibinfo {author} {\bibfnamefont
  {W.}~\bibnamefont {N\"ortersh\"auser}}, \bibinfo {author} {\bibfnamefont
  {T.}~\bibnamefont {Papenbrock}}, \bibinfo {author} {\bibfnamefont
  {J.}~\bibnamefont {Papuga}}, \bibinfo {author} {\bibfnamefont
  {A.}~\bibnamefont {Schwenk}}, \bibinfo {author} {\bibfnamefont
  {J.}~\bibnamefont {Simonis}}, \bibinfo {author} {\bibfnamefont {K.~A.}\
  \bibnamefont {Wendt}}, \ and\ \bibinfo {author} {\bibfnamefont {D.~T.}\
  \bibnamefont {Yordanov}},\ }\href {\doibase 10.1038/nphys3645} {\bibfield
  {journal} {\bibinfo  {journal} {Nat. Phys.}\ }\textbf {\bibinfo {volume}
  {12}},\ \bibinfo {pages} {594} (\bibinfo {year} {2016})}\BibitemShut
  {NoStop}%
\bibitem [{\citenamefont {Miller}\ \emph {et~al.}(2019)\citenamefont {Miller},
  \citenamefont {Minamisono}, \citenamefont {Klose}, \citenamefont {Garand},
  \citenamefont {Kujawa}, \citenamefont {Lantis}, \citenamefont {Liu},
  \citenamefont {Maa{\ss}}, \citenamefont {Mantica}, \citenamefont
  {Nazarewicz}, \citenamefont {N\"ortersh\"auser}, \citenamefont {Pineda},
  \citenamefont {Reinhard}, \citenamefont {Rossi}, \citenamefont {Sommer},
  \citenamefont {Sumithrarachchi}, \citenamefont {Teigelh\"ofer},\ and\
  \citenamefont {Watkins}}]{Miller2019_NatPhys15-432}%
  \BibitemOpen
  \bibfield  {author} {\bibinfo {author} {\bibfnamefont {A.~J.}\ \bibnamefont
  {Miller}}, \bibinfo {author} {\bibfnamefont {K.}~\bibnamefont {Minamisono}},
  \bibinfo {author} {\bibfnamefont {A.}~\bibnamefont {Klose}}, \bibinfo
  {author} {\bibfnamefont {D.}~\bibnamefont {Garand}}, \bibinfo {author}
  {\bibfnamefont {C.}~\bibnamefont {Kujawa}}, \bibinfo {author} {\bibfnamefont
  {J.~D.}\ \bibnamefont {Lantis}}, \bibinfo {author} {\bibfnamefont
  {Y.}~\bibnamefont {Liu}}, \bibinfo {author} {\bibfnamefont {B.}~\bibnamefont
  {Maa{\ss}}}, \bibinfo {author} {\bibfnamefont {P.~F.}\ \bibnamefont
  {Mantica}}, \bibinfo {author} {\bibfnamefont {W.}~\bibnamefont {Nazarewicz}},
  \bibinfo {author} {\bibfnamefont {W.}~\bibnamefont {N\"ortersh\"auser}},
  \bibinfo {author} {\bibfnamefont {S.~V.}\ \bibnamefont {Pineda}}, \bibinfo
  {author} {\bibfnamefont {P.-G.}\ \bibnamefont {Reinhard}}, \bibinfo {author}
  {\bibfnamefont {D.~M.}\ \bibnamefont {Rossi}}, \bibinfo {author}
  {\bibfnamefont {F.}~\bibnamefont {Sommer}}, \bibinfo {author} {\bibfnamefont
  {C.}~\bibnamefont {Sumithrarachchi}}, \bibinfo {author} {\bibfnamefont
  {A.}~\bibnamefont {Teigelh\"ofer}}, \ and\ \bibinfo {author} {\bibfnamefont
  {J.}~\bibnamefont {Watkins}},\ }\href {\doibase 10.1038/s41567-019-0416-9}
  {\bibfield  {journal} {\bibinfo  {journal} {Nat. Phys.}\ }\textbf {\bibinfo
  {volume} {15}},\ \bibinfo {pages} {432} (\bibinfo {year} {2019})}\BibitemShut
  {NoStop}%
\bibitem [{\citenamefont {de~Groote}\ \emph {et~al.}(2020)\citenamefont
  {de~Groote}, \citenamefont {P.}, \citenamefont {Billowes}, \citenamefont
  {Binnersley}, \citenamefont {Bissell}, \citenamefont {Cocolios},
  \citenamefont {Day~Goodacre}, \citenamefont {Farooq-Smith}, \citenamefont
  {Fedorov}, \citenamefont {Flanagan}, \citenamefont {Franchoo}, \citenamefont
  {Garcia~Ruiz}, \citenamefont {Gins}, \citenamefont {Holt}, \citenamefont
  {Koszor\'us}, \citenamefont {Lynch}, \citenamefont {Miyagi}, \citenamefont
  {Nazarewicz}, \citenamefont {Neyens}, \citenamefont {Reinhard}, \citenamefont
  {Rothe}, \citenamefont {Stroke}, \citenamefont {Vernon}, \citenamefont
  {Wendt}, \citenamefont {Wilkins}, \citenamefont {Xu},\ and\ \citenamefont
  {Yang}}]{Groote2020_NatPhys16-620}%
  \BibitemOpen
  \bibfield  {author} {\bibinfo {author} {\bibnamefont {de~Groote}}, \bibinfo
  {author} {\bibfnamefont {R.}~\bibnamefont {P.}}, \bibinfo {author}
  {\bibfnamefont {J.}~\bibnamefont {Billowes}}, \bibinfo {author}
  {\bibfnamefont {C.~L.}\ \bibnamefont {Binnersley}}, \bibinfo {author}
  {\bibfnamefont {M.~L.}\ \bibnamefont {Bissell}}, \bibinfo {author}
  {\bibfnamefont {T.~E.}\ \bibnamefont {Cocolios}}, \bibinfo {author}
  {\bibfnamefont {T.}~\bibnamefont {Day~Goodacre}}, \bibinfo {author}
  {\bibfnamefont {G.~J.}\ \bibnamefont {Farooq-Smith}}, \bibinfo {author}
  {\bibfnamefont {D.~V.}\ \bibnamefont {Fedorov}}, \bibinfo {author}
  {\bibfnamefont {K.~T.}\ \bibnamefont {Flanagan}}, \bibinfo {author}
  {\bibfnamefont {S.}~\bibnamefont {Franchoo}}, \bibinfo {author}
  {\bibfnamefont {R.~F.}\ \bibnamefont {Garcia~Ruiz}}, \bibinfo {author}
  {\bibfnamefont {W.}~\bibnamefont {Gins}}, \bibinfo {author} {\bibfnamefont
  {J.~D.}\ \bibnamefont {Holt}}, \bibinfo {author} {\bibfnamefont
  {A.}~\bibnamefont {Koszor\'us}}, \bibinfo {author} {\bibfnamefont {K.~M.}\
  \bibnamefont {Lynch}}, \bibinfo {author} {\bibfnamefont {T.}~\bibnamefont
  {Miyagi}}, \bibinfo {author} {\bibfnamefont {W.}~\bibnamefont {Nazarewicz}},
  \bibinfo {author} {\bibfnamefont {G.}~\bibnamefont {Neyens}}, \bibinfo
  {author} {\bibfnamefont {P.-G.}\ \bibnamefont {Reinhard}}, \bibinfo {author}
  {\bibfnamefont {S.}~\bibnamefont {Rothe}}, \bibinfo {author} {\bibfnamefont
  {H.~H.}\ \bibnamefont {Stroke}}, \bibinfo {author} {\bibfnamefont {A.~R.}\
  \bibnamefont {Vernon}}, \bibinfo {author} {\bibfnamefont {K.~D.~A.}\
  \bibnamefont {Wendt}}, \bibinfo {author} {\bibfnamefont {S.~G.}\ \bibnamefont
  {Wilkins}}, \bibinfo {author} {\bibfnamefont {Z.~Y.}\ \bibnamefont {Xu}}, \
  and\ \bibinfo {author} {\bibfnamefont {X.~F.}\ \bibnamefont {Yang}},\ }\href
  {\doibase 10.1038/s41567-020-0868-y} {\bibfield  {journal} {\bibinfo
  {journal} {Nat. Phys.}\ }\textbf {\bibinfo {volume} {16}},\ \bibinfo {pages}
  {620} (\bibinfo {year} {2020})}\BibitemShut {NoStop}%
\bibitem [{\citenamefont {Koszor\'us}\ \emph {et~al.}(2021)\citenamefont
  {Koszor\'us}, \citenamefont {Yang}, \citenamefont {Jiang}, \citenamefont
  {Novario}, \citenamefont {Bai}, \citenamefont {Billowes}, \citenamefont
  {Binnersley}, \citenamefont {Bissell}, \citenamefont {Cocolios},
  \citenamefont {Cooper}, \citenamefont {de~Groote}, \citenamefont {Ekstr\"om},
  \citenamefont {Flanagan}, \citenamefont {Forss\'en}, \citenamefont
  {Franchoo}, \citenamefont {Ruiz}, \citenamefont {Gustafsson}, \citenamefont
  {Hagen}, \citenamefont {Jansen}, \citenamefont {Kanellakopoulos},
  \citenamefont {Kortelainen}, \citenamefont {Nazarewicz}, \citenamefont
  {Neyens}, \citenamefont {Papenbrock}, \citenamefont {Reinhard}, \citenamefont
  {Ricketts}, \citenamefont {Sahoo}, \citenamefont {Vernon},\ and\
  \citenamefont {Wilkins}}]{Koszorus2021_NatPhys17-439}%
  \BibitemOpen
  \bibfield  {author} {\bibinfo {author} {\bibfnamefont {A.}~\bibnamefont
  {Koszor\'us}}, \bibinfo {author} {\bibfnamefont {X.~F.}\ \bibnamefont
  {Yang}}, \bibinfo {author} {\bibfnamefont {W.~G.}\ \bibnamefont {Jiang}},
  \bibinfo {author} {\bibfnamefont {S.~J.}\ \bibnamefont {Novario}}, \bibinfo
  {author} {\bibfnamefont {S.~W.}\ \bibnamefont {Bai}}, \bibinfo {author}
  {\bibfnamefont {J.}~\bibnamefont {Billowes}}, \bibinfo {author}
  {\bibfnamefont {C.~L.}\ \bibnamefont {Binnersley}}, \bibinfo {author}
  {\bibfnamefont {M.~L.}\ \bibnamefont {Bissell}}, \bibinfo {author}
  {\bibfnamefont {T.~E.}\ \bibnamefont {Cocolios}}, \bibinfo {author}
  {\bibfnamefont {B.~S.}\ \bibnamefont {Cooper}}, \bibinfo {author}
  {\bibfnamefont {R.~P.}\ \bibnamefont {de~Groote}}, \bibinfo {author}
  {\bibfnamefont {A.}~\bibnamefont {Ekstr\"om}}, \bibinfo {author}
  {\bibfnamefont {K.~T.}\ \bibnamefont {Flanagan}}, \bibinfo {author}
  {\bibfnamefont {C.}~\bibnamefont {Forss\'en}}, \bibinfo {author}
  {\bibfnamefont {S.}~\bibnamefont {Franchoo}}, \bibinfo {author}
  {\bibfnamefont {R.~F.~G.}\ \bibnamefont {Ruiz}}, \bibinfo {author}
  {\bibfnamefont {F.~P.}\ \bibnamefont {Gustafsson}}, \bibinfo {author}
  {\bibfnamefont {G.}~\bibnamefont {Hagen}}, \bibinfo {author} {\bibfnamefont
  {G.~R.}\ \bibnamefont {Jansen}}, \bibinfo {author} {\bibfnamefont
  {A.}~\bibnamefont {Kanellakopoulos}}, \bibinfo {author} {\bibfnamefont
  {M.}~\bibnamefont {Kortelainen}}, \bibinfo {author} {\bibfnamefont
  {W.}~\bibnamefont {Nazarewicz}}, \bibinfo {author} {\bibfnamefont
  {G.}~\bibnamefont {Neyens}}, \bibinfo {author} {\bibfnamefont
  {T.}~\bibnamefont {Papenbrock}}, \bibinfo {author} {\bibfnamefont {P.-G.}\
  \bibnamefont {Reinhard}}, \bibinfo {author} {\bibfnamefont {C.~M.}\
  \bibnamefont {Ricketts}}, \bibinfo {author} {\bibfnamefont {B.~K.}\
  \bibnamefont {Sahoo}}, \bibinfo {author} {\bibfnamefont {A.~R.}\ \bibnamefont
  {Vernon}}, \ and\ \bibinfo {author} {\bibfnamefont {S.~G.}\ \bibnamefont
  {Wilkins}},\ }\href {\doibase 10.1038/s41567-020-01136-5} {\bibfield
  {journal} {\bibinfo  {journal} {Nat. Phys.}\ }\textbf {\bibinfo {volume}
  {17}},\ \bibinfo {pages} {439} (\bibinfo {year} {2021})}\BibitemShut
  {NoStop}%
\bibitem [{\citenamefont {Bohr}\ and\ \citenamefont
  {Mottelson}(1969)}]{Bohr1969_Book}%
  \BibitemOpen
  \bibfield  {author} {\bibinfo {author} {\bibfnamefont {A.}~\bibnamefont
  {Bohr}}\ and\ \bibinfo {author} {\bibfnamefont {B.~R.}\ \bibnamefont
  {Mottelson}},\ }\href@noop {} {\emph {\bibinfo {title} {Nuclear Structure,
  Vol. I Single-particle Motion}}}\ (\bibinfo  {publisher} {Benjamin},\
  \bibinfo {year} {1969})\BibitemShut {NoStop}%
\bibitem [{\citenamefont {Zeng}(1957)}]{Zeng1957_ActaPhysSin13-357}%
  \BibitemOpen
  \bibfield  {author} {\bibinfo {author} {\bibfnamefont {J.~Y.}\ \bibnamefont
  {Zeng}},\ }\href@noop {} {\bibfield  {journal} {\bibinfo  {journal} {Acta
  Phys. Sin.}\ }\textbf {\bibinfo {volume} {13}},\ \bibinfo {pages} {357}
  (\bibinfo {year} {1957})}\BibitemShut {NoStop}%
\bibitem [{\citenamefont {Nerlo-Pomorska}\ and\ \citenamefont
  {Pomorski}(1993)}]{Nerlo-Pomorska1993_ZPA344-359}%
  \BibitemOpen
  \bibfield  {author} {\bibinfo {author} {\bibfnamefont {B.}~\bibnamefont
  {Nerlo-Pomorska}}\ and\ \bibinfo {author} {\bibfnamefont {K.}~\bibnamefont
  {Pomorski}},\ }\href {\doibase 10.1007/BF01283190} {\bibfield  {journal}
  {\bibinfo  {journal} {Z. Phys. A}\ }\textbf {\bibinfo {volume} {344}},\
  \bibinfo {pages} {359} (\bibinfo {year} {1993})}\BibitemShut {NoStop}%
\bibitem [{\citenamefont {Duflo}(1994)}]{Duflo1994_NPA576-29}%
  \BibitemOpen
  \bibfield  {author} {\bibinfo {author} {\bibfnamefont {J.}~\bibnamefont
  {Duflo}},\ }\href {\doibase 10.1016/0375-9474(94)90737-4} {\bibfield
  {journal} {\bibinfo  {journal} {Nucl. Phys. A}\ }\textbf {\bibinfo {volume}
  {576}},\ \bibinfo {pages} {29} (\bibinfo {year} {1994})}\BibitemShut
  {NoStop}%
\bibitem [{\citenamefont {Zhang}\ \emph {et~al.}(2002)\citenamefont {Zhang},
  \citenamefont {Meng}, \citenamefont {Zhou},\ and\ \citenamefont
  {Zeng}}]{Zhang2002_EPJA13-285}%
  \BibitemOpen
  \bibfield  {author} {\bibinfo {author} {\bibfnamefont {S.}~\bibnamefont
  {Zhang}}, \bibinfo {author} {\bibfnamefont {J.}~\bibnamefont {Meng}},
  \bibinfo {author} {\bibfnamefont {S.-G.}\ \bibnamefont {Zhou}}, \ and\
  \bibinfo {author} {\bibfnamefont {J.}~\bibnamefont {Zeng}},\ }\href {\doibase
  10.1007/s10050-002-8757-6} {\bibfield  {journal} {\bibinfo  {journal} {Eur.
  Phys. J. A}\ }\textbf {\bibinfo {volume} {13}},\ \bibinfo {pages} {285}
  (\bibinfo {year} {2002})}\BibitemShut {NoStop}%
\bibitem [{\citenamefont {Lei}\ \emph {et~al.}(2009)\citenamefont {Lei},
  \citenamefont {Zhang},\ and\ \citenamefont {Zeng}}]{Lei2009_CTP51-123}%
  \BibitemOpen
  \bibfield  {author} {\bibinfo {author} {\bibfnamefont {Y.-A.}\ \bibnamefont
  {Lei}}, \bibinfo {author} {\bibfnamefont {Z.-H.}\ \bibnamefont {Zhang}}, \
  and\ \bibinfo {author} {\bibfnamefont {J.-Y.}\ \bibnamefont {Zeng}},\ }\href
  {\doibase 10.1088/0253-6102/51/1/23} {\bibfield  {journal} {\bibinfo
  {journal} {Commun. Theor. Phys.}\ }\textbf {\bibinfo {volume} {51}},\
  \bibinfo {pages} {123} (\bibinfo {year} {2009})}\BibitemShut {NoStop}%
\bibitem [{\citenamefont {Wang}\ and\ \citenamefont
  {Li}(2013)}]{Wang2013_PRC88-011301}%
  \BibitemOpen
  \bibfield  {author} {\bibinfo {author} {\bibfnamefont {N.}~\bibnamefont
  {Wang}}\ and\ \bibinfo {author} {\bibfnamefont {T.}~\bibnamefont {Li}},\
  }\href {\doibase 10.1103/PhysRevC.88.011301} {\bibfield  {journal} {\bibinfo
  {journal} {Phys. Rev. C}\ }\textbf {\bibinfo {volume} {88}},\ \bibinfo
  {pages} {011301(R)} (\bibinfo {year} {2013})}\BibitemShut {NoStop}%
\bibitem [{\citenamefont {Bayram}\ \emph {et~al.}(2013)\citenamefont {Bayram},
  \citenamefont {Akkoyun}, \citenamefont {Kara},\ and\ \citenamefont
  {Sinan}}]{Bayram2013_APPB44-1791}%
  \BibitemOpen
  \bibfield  {author} {\bibinfo {author} {\bibfnamefont {T.}~\bibnamefont
  {Bayram}}, \bibinfo {author} {\bibfnamefont {S.}~\bibnamefont {Akkoyun}},
  \bibinfo {author} {\bibfnamefont {S.}~\bibnamefont {Kara}}, \ and\ \bibinfo
  {author} {\bibfnamefont {A.}~\bibnamefont {Sinan}},\ }\href {\doibase
  10.5506/APhysPolB.44.1791} {\bibfield  {journal} {\bibinfo  {journal} {Acta
  Phys. Pol. B}\ }\textbf {\bibinfo {volume} {44}},\ \bibinfo {pages} {1791}
  (\bibinfo {year} {2013})}\BibitemShut {NoStop}%
\bibitem [{\citenamefont {Buchinger}\ \emph {et~al.}(1994)\citenamefont
  {Buchinger}, \citenamefont {Crawford}, \citenamefont {Dutta}, \citenamefont
  {Pearson},\ and\ \citenamefont {Tondeur}}]{Buchinger1994_PRC49-1402}%
  \BibitemOpen
  \bibfield  {author} {\bibinfo {author} {\bibfnamefont {F.}~\bibnamefont
  {Buchinger}}, \bibinfo {author} {\bibfnamefont {J.~E.}\ \bibnamefont
  {Crawford}}, \bibinfo {author} {\bibfnamefont {A.~K.}\ \bibnamefont {Dutta}},
  \bibinfo {author} {\bibfnamefont {J.~M.}\ \bibnamefont {Pearson}}, \ and\
  \bibinfo {author} {\bibfnamefont {F.}~\bibnamefont {Tondeur}},\ }\href
  {\doibase 10.1103/PhysRevC.49.1402} {\bibfield  {journal} {\bibinfo
  {journal} {Phys. Rev. C}\ }\textbf {\bibinfo {volume} {49}},\ \bibinfo
  {pages} {1402} (\bibinfo {year} {1994})}\BibitemShut {NoStop}%
\bibitem [{\citenamefont {Buchinger}\ \emph {et~al.}(2001)\citenamefont
  {Buchinger}, \citenamefont {Pearson},\ and\ \citenamefont
  {Goriely}}]{Buchinger2001_PRC64-067303}%
  \BibitemOpen
  \bibfield  {author} {\bibinfo {author} {\bibfnamefont {F.}~\bibnamefont
  {Buchinger}}, \bibinfo {author} {\bibfnamefont {J.~M.}\ \bibnamefont
  {Pearson}}, \ and\ \bibinfo {author} {\bibfnamefont {S.}~\bibnamefont
  {Goriely}},\ }\href {\doibase 10.1103/PhysRevC.64.067303} {\bibfield
  {journal} {\bibinfo  {journal} {Phys. Rev. C}\ }\textbf {\bibinfo {volume}
  {64}},\ \bibinfo {pages} {067303} (\bibinfo {year} {2001})}\BibitemShut
  {NoStop}%
\bibitem [{\citenamefont {Buchinger}\ and\ \citenamefont
  {Pearson}(2005)}]{Buchinger2005_PRC72-057305}%
  \BibitemOpen
  \bibfield  {author} {\bibinfo {author} {\bibfnamefont {F.}~\bibnamefont
  {Buchinger}}\ and\ \bibinfo {author} {\bibfnamefont {J.~M.}\ \bibnamefont
  {Pearson}},\ }\href {\doibase 10.1103/PhysRevC.72.057305} {\bibfield
  {journal} {\bibinfo  {journal} {Phys. Rev. C}\ }\textbf {\bibinfo {volume}
  {72}},\ \bibinfo {pages} {057305} (\bibinfo {year} {2005})}\BibitemShut
  {NoStop}%
\bibitem [{\citenamefont {Iimura}\ and\ \citenamefont
  {Buchinger}(2008)}]{Iimura2008_PRC78-067301}%
  \BibitemOpen
  \bibfield  {author} {\bibinfo {author} {\bibfnamefont {H.}~\bibnamefont
  {Iimura}}\ and\ \bibinfo {author} {\bibfnamefont {F.}~\bibnamefont
  {Buchinger}},\ }\href {\doibase 10.1103/PhysRevC.78.067301} {\bibfield
  {journal} {\bibinfo  {journal} {Phys. Rev. C}\ }\textbf {\bibinfo {volume}
  {78}},\ \bibinfo {pages} {067301} (\bibinfo {year} {2008})}\BibitemShut
  {NoStop}%
\bibitem [{\citenamefont {Lalazissis}\ \emph {et~al.}(1999)\citenamefont
  {Lalazissis}, \citenamefont {Raman},\ and\ \citenamefont
  {Ring}}]{Lalazissis1999_ADNDT71-1}%
  \BibitemOpen
  \bibfield  {author} {\bibinfo {author} {\bibfnamefont {G.~A.}\ \bibnamefont
  {Lalazissis}}, \bibinfo {author} {\bibfnamefont {S.}~\bibnamefont {Raman}}, \
  and\ \bibinfo {author} {\bibfnamefont {P.}~\bibnamefont {Ring}},\ }\href
  {\doibase 10.1006/adnd.1998.0795} {\bibfield  {journal} {\bibinfo  {journal}
  {At. Data Nucl. Data Tables}\ }\textbf {\bibinfo {volume} {71}},\ \bibinfo
  {pages} {1} (\bibinfo {year} {1999})}\BibitemShut {NoStop}%
\bibitem [{\citenamefont {Geng}\ \emph {et~al.}(2005)\citenamefont {Geng},
  \citenamefont {Toki},\ and\ \citenamefont {Meng}}]{Geng2005_PTP113-785}%
  \BibitemOpen
  \bibfield  {author} {\bibinfo {author} {\bibfnamefont {L.~S.}\ \bibnamefont
  {Geng}}, \bibinfo {author} {\bibfnamefont {H.}~\bibnamefont {Toki}}, \ and\
  \bibinfo {author} {\bibfnamefont {J.}~\bibnamefont {Meng}},\ }\href {\doibase
  10.1143/PTP.113.785} {\bibfield  {journal} {\bibinfo  {journal} {Prog. Theo.
  Phys.}\ }\textbf {\bibinfo {volume} {113}},\ \bibinfo {pages} {785} (\bibinfo
  {year} {2005})}\BibitemShut {NoStop}%
\bibitem [{\citenamefont {Zhao}\ \emph {et~al.}(2010)\citenamefont {Zhao},
  \citenamefont {Li}, \citenamefont {Yao},\ and\ \citenamefont
  {Meng}}]{Zhao2010_PRC82-054319}%
  \BibitemOpen
  \bibfield  {author} {\bibinfo {author} {\bibfnamefont {P.~W.}\ \bibnamefont
  {Zhao}}, \bibinfo {author} {\bibfnamefont {Z.~P.}\ \bibnamefont {Li}},
  \bibinfo {author} {\bibfnamefont {J.~M.}\ \bibnamefont {Yao}}, \ and\
  \bibinfo {author} {\bibfnamefont {J.}~\bibnamefont {Meng}},\ }\href {\doibase
  10.1103/PhysRevC.82.054319} {\bibfield  {journal} {\bibinfo  {journal} {Phys.
  Rev. C}\ }\textbf {\bibinfo {volume} {82}},\ \bibinfo {pages} {054319}
  (\bibinfo {year} {2010})}\BibitemShut {NoStop}%
\bibitem [{\citenamefont {Xia}\ \emph {et~al.}(2018)\citenamefont {Xia},
  \citenamefont {Lim}, \citenamefont {Zhao}, \citenamefont {Liang},
  \citenamefont {Qu}, \citenamefont {Chen}, \citenamefont {Liu}, \citenamefont
  {Zhang}, \citenamefont {Zhang}, \citenamefont {Kim},\ and\ \citenamefont
  {Meng}}]{Xia2018_ADNDT121-122-1}%
  \BibitemOpen
  \bibfield  {author} {\bibinfo {author} {\bibfnamefont {X.~W.}\ \bibnamefont
  {Xia}}, \bibinfo {author} {\bibfnamefont {Y.}~\bibnamefont {Lim}}, \bibinfo
  {author} {\bibfnamefont {P.~W.}\ \bibnamefont {Zhao}}, \bibinfo {author}
  {\bibfnamefont {H.~Z.}\ \bibnamefont {Liang}}, \bibinfo {author}
  {\bibfnamefont {X.~Y.}\ \bibnamefont {Qu}}, \bibinfo {author} {\bibfnamefont
  {Y.}~\bibnamefont {Chen}}, \bibinfo {author} {\bibfnamefont {H.}~\bibnamefont
  {Liu}}, \bibinfo {author} {\bibfnamefont {L.~F.}\ \bibnamefont {Zhang}},
  \bibinfo {author} {\bibfnamefont {S.~Q.}\ \bibnamefont {Zhang}}, \bibinfo
  {author} {\bibfnamefont {Y.}~\bibnamefont {Kim}}, \ and\ \bibinfo {author}
  {\bibfnamefont {J.}~\bibnamefont {Meng}},\ }\href {\doibase
  10.1016/j.adt.2017.09.001} {\bibfield  {journal} {\bibinfo  {journal} {At.
  Data Nucl. Data Tables}\ }\textbf {\bibinfo {volume} {121-122}},\ \bibinfo
  {pages} {1} (\bibinfo {year} {2018})}\BibitemShut {NoStop}%
\bibitem [{\citenamefont {Zhang}\ \emph {et~al.}(2020)\citenamefont {Zhang},
  \citenamefont {Cheoun}, \citenamefont {Choi}, \citenamefont {Chong},
  \citenamefont {Dong}, \citenamefont {Geng}, \citenamefont {Ha}, \citenamefont
  {He}, \citenamefont {Heo}, \citenamefont {Ho}, \citenamefont {In},
  \citenamefont {Kim}, \citenamefont {Kim}, \citenamefont {Lee}, \citenamefont
  {Lee}, \citenamefont {Li}, \citenamefont {Luo}, \citenamefont {Meng},
  \citenamefont {Mun}, \citenamefont {Niu}, \citenamefont {Pan}, \citenamefont
  {Papakonstantinou}, \citenamefont {Shang}, \citenamefont {Shen},
  \citenamefont {Shen}, \citenamefont {Sun}, \citenamefont {Sun}, \citenamefont
  {Tam}, \citenamefont {Thaivayongnou}, \citenamefont {Wang}, \citenamefont
  {Wong}, \citenamefont {Xia}, \citenamefont {Yan}, \citenamefont {Yeung},
  \citenamefont {Yiu}, \citenamefont {Zhang}, \citenamefont {Zhang},\ and\
  \citenamefont {Zhou}}]{Zhang2020_PRC102-024314}%
  \BibitemOpen
  \bibfield  {author} {\bibinfo {author} {\bibfnamefont {K.}~\bibnamefont
  {Zhang}}, \bibinfo {author} {\bibfnamefont {M.-K.}\ \bibnamefont {Cheoun}},
  \bibinfo {author} {\bibfnamefont {Y.-B.}\ \bibnamefont {Choi}}, \bibinfo
  {author} {\bibfnamefont {P.~S.}\ \bibnamefont {Chong}}, \bibinfo {author}
  {\bibfnamefont {J.}~\bibnamefont {Dong}}, \bibinfo {author} {\bibfnamefont
  {L.}~\bibnamefont {Geng}}, \bibinfo {author} {\bibfnamefont {E.}~\bibnamefont
  {Ha}}, \bibinfo {author} {\bibfnamefont {X.}~\bibnamefont {He}}, \bibinfo
  {author} {\bibfnamefont {C.}~\bibnamefont {Heo}}, \bibinfo {author}
  {\bibfnamefont {M.~C.}\ \bibnamefont {Ho}}, \bibinfo {author} {\bibfnamefont
  {E.~J.}\ \bibnamefont {In}}, \bibinfo {author} {\bibfnamefont
  {S.}~\bibnamefont {Kim}}, \bibinfo {author} {\bibfnamefont {Y.}~\bibnamefont
  {Kim}}, \bibinfo {author} {\bibfnamefont {C.-H.}\ \bibnamefont {Lee}},
  \bibinfo {author} {\bibfnamefont {J.}~\bibnamefont {Lee}}, \bibinfo {author}
  {\bibfnamefont {Z.}~\bibnamefont {Li}}, \bibinfo {author} {\bibfnamefont
  {T.}~\bibnamefont {Luo}}, \bibinfo {author} {\bibfnamefont {J.}~\bibnamefont
  {Meng}}, \bibinfo {author} {\bibfnamefont {M.-H.}\ \bibnamefont {Mun}},
  \bibinfo {author} {\bibfnamefont {Z.}~\bibnamefont {Niu}}, \bibinfo {author}
  {\bibfnamefont {C.}~\bibnamefont {Pan}}, \bibinfo {author} {\bibfnamefont
  {P.}~\bibnamefont {Papakonstantinou}}, \bibinfo {author} {\bibfnamefont
  {X.}~\bibnamefont {Shang}}, \bibinfo {author} {\bibfnamefont
  {C.}~\bibnamefont {Shen}}, \bibinfo {author} {\bibfnamefont {G.}~\bibnamefont
  {Shen}}, \bibinfo {author} {\bibfnamefont {W.}~\bibnamefont {Sun}}, \bibinfo
  {author} {\bibfnamefont {X.-X.}\ \bibnamefont {Sun}}, \bibinfo {author}
  {\bibfnamefont {C.~K.}\ \bibnamefont {Tam}}, \bibinfo {author} {\bibnamefont
  {Thaivayongnou}}, \bibinfo {author} {\bibfnamefont {C.}~\bibnamefont {Wang}},
  \bibinfo {author} {\bibfnamefont {S.~H.}\ \bibnamefont {Wong}}, \bibinfo
  {author} {\bibfnamefont {X.}~\bibnamefont {Xia}}, \bibinfo {author}
  {\bibfnamefont {Y.}~\bibnamefont {Yan}}, \bibinfo {author} {\bibfnamefont
  {R.~W.-Y.}\ \bibnamefont {Yeung}}, \bibinfo {author} {\bibfnamefont {T.~C.}\
  \bibnamefont {Yiu}}, \bibinfo {author} {\bibfnamefont {S.}~\bibnamefont
  {Zhang}}, \bibinfo {author} {\bibfnamefont {W.}~\bibnamefont {Zhang}}, \ and\
  \bibinfo {author} {\bibfnamefont {S.-G.}\ \bibnamefont {Zhou}} (\bibinfo
  {collaboration} {DRHBc Mass Table Collaboration}),\ }\href {\doibase
  10.1103/PhysRevC.102.024314} {\bibfield  {journal} {\bibinfo  {journal}
  {Phys. Rev. C}\ }\textbf {\bibinfo {volume} {102}},\ \bibinfo {pages}
  {024314} (\bibinfo {year} {2020})}\BibitemShut {NoStop}%
\bibitem [{\citenamefont {An}\ \emph {et~al.}(2020)\citenamefont {An},
  \citenamefont {Geng},\ and\ \citenamefont {Zhang}}]{An2020_PRC102-024307}%
  \BibitemOpen
  \bibfield  {author} {\bibinfo {author} {\bibfnamefont {R.}~\bibnamefont
  {An}}, \bibinfo {author} {\bibfnamefont {L.-S.}\ \bibnamefont {Geng}}, \ and\
  \bibinfo {author} {\bibfnamefont {S.-S.}\ \bibnamefont {Zhang}},\ }\href
  {\doibase 10.1103/PhysRevC.102.024307} {\bibfield  {journal} {\bibinfo
  {journal} {Phys. Rev. C}\ }\textbf {\bibinfo {volume} {102}},\ \bibinfo
  {pages} {024307} (\bibinfo {year} {2020})}\BibitemShut {NoStop}%
\bibitem [{\citenamefont {Perera}\ \emph {et~al.}(2021)\citenamefont {Perera},
  \citenamefont {Afanasjev},\ and\ \citenamefont
  {Ring}}]{Perera2021_PRC104-064313}%
  \BibitemOpen
  \bibfield  {author} {\bibinfo {author} {\bibfnamefont {U.~C.}\ \bibnamefont
  {Perera}}, \bibinfo {author} {\bibfnamefont {A.~V.}\ \bibnamefont
  {Afanasjev}}, \ and\ \bibinfo {author} {\bibfnamefont {P.}~\bibnamefont
  {Ring}},\ }\href {\doibase 10.1103/PhysRevC.104.064313} {\bibfield  {journal}
  {\bibinfo  {journal} {Phys. Rev. C}\ }\textbf {\bibinfo {volume} {104}},\
  \bibinfo {pages} {064313} (\bibinfo {year} {2021})}\BibitemShut {NoStop}%
\bibitem [{\citenamefont {Zhang}\ \emph {et~al.}(2022)\citenamefont {Zhang},
  \citenamefont {Cheoun}, \citenamefont {Choi}, \citenamefont {Chong},
  \citenamefont {Dong}, \citenamefont {Dong}, \citenamefont {Du}, \citenamefont
  {Geng}, \citenamefont {Ha}, \citenamefont {He}, \citenamefont {Heo},
  \citenamefont {Ho}, \citenamefont {In}, \citenamefont {Kim}, \citenamefont
  {Kim}, \citenamefont {Lee}, \citenamefont {Lee}, \citenamefont {Li},
  \citenamefont {Li}, \citenamefont {Luo}, \citenamefont {Meng}, \citenamefont
  {Mun}, \citenamefont {Niu}, \citenamefont {Pan}, \citenamefont
  {Papakonstantinou}, \citenamefont {Shang}, \citenamefont {Shen},
  \citenamefont {Shen}, \citenamefont {Sun}, \citenamefont {Sun}, \citenamefont
  {Tam}, \citenamefont {Thaivayongnou}, \citenamefont {Wang}, \citenamefont
  {Wang}, \citenamefont {Wong}, \citenamefont {Wu}, \citenamefont {Wu},
  \citenamefont {Xia}, \citenamefont {Yan}, \citenamefont {Yeung},
  \citenamefont {Yiu}, \citenamefont {Zhang}, \citenamefont {Zhang},
  \citenamefont {Zhang}, \citenamefont {Zhao},\ and\ \citenamefont
  {Zhou}}]{Zhang2022_ADNDT144-101488}%
  \BibitemOpen
  \bibfield  {author} {\bibinfo {author} {\bibfnamefont {K.}~\bibnamefont
  {Zhang}}, \bibinfo {author} {\bibfnamefont {M.-K.}\ \bibnamefont {Cheoun}},
  \bibinfo {author} {\bibfnamefont {Y.-B.}\ \bibnamefont {Choi}}, \bibinfo
  {author} {\bibfnamefont {P.~S.}\ \bibnamefont {Chong}}, \bibinfo {author}
  {\bibfnamefont {J.}~\bibnamefont {Dong}}, \bibinfo {author} {\bibfnamefont
  {Z.}~\bibnamefont {Dong}}, \bibinfo {author} {\bibfnamefont {X.}~\bibnamefont
  {Du}}, \bibinfo {author} {\bibfnamefont {L.}~\bibnamefont {Geng}}, \bibinfo
  {author} {\bibfnamefont {E.}~\bibnamefont {Ha}}, \bibinfo {author}
  {\bibfnamefont {X.-T.}\ \bibnamefont {He}}, \bibinfo {author} {\bibfnamefont
  {C.}~\bibnamefont {Heo}}, \bibinfo {author} {\bibfnamefont {M.~C.}\
  \bibnamefont {Ho}}, \bibinfo {author} {\bibfnamefont {E.~J.}\ \bibnamefont
  {In}}, \bibinfo {author} {\bibfnamefont {S.}~\bibnamefont {Kim}}, \bibinfo
  {author} {\bibfnamefont {Y.}~\bibnamefont {Kim}}, \bibinfo {author}
  {\bibfnamefont {C.-H.}\ \bibnamefont {Lee}}, \bibinfo {author} {\bibfnamefont
  {J.}~\bibnamefont {Lee}}, \bibinfo {author} {\bibfnamefont {H.}~\bibnamefont
  {Li}}, \bibinfo {author} {\bibfnamefont {Z.}~\bibnamefont {Li}}, \bibinfo
  {author} {\bibfnamefont {T.}~\bibnamefont {Luo}}, \bibinfo {author}
  {\bibfnamefont {J.}~\bibnamefont {Meng}}, \bibinfo {author} {\bibfnamefont
  {M.-H.}\ \bibnamefont {Mun}}, \bibinfo {author} {\bibfnamefont
  {Z.}~\bibnamefont {Niu}}, \bibinfo {author} {\bibfnamefont {C.}~\bibnamefont
  {Pan}}, \bibinfo {author} {\bibfnamefont {P.}~\bibnamefont
  {Papakonstantinou}}, \bibinfo {author} {\bibfnamefont {X.}~\bibnamefont
  {Shang}}, \bibinfo {author} {\bibfnamefont {C.}~\bibnamefont {Shen}},
  \bibinfo {author} {\bibfnamefont {G.}~\bibnamefont {Shen}}, \bibinfo {author}
  {\bibfnamefont {W.}~\bibnamefont {Sun}}, \bibinfo {author} {\bibfnamefont
  {X.-X.}\ \bibnamefont {Sun}}, \bibinfo {author} {\bibfnamefont {C.~K.}\
  \bibnamefont {Tam}}, \bibinfo {author} {\bibnamefont {Thaivayongnou}},
  \bibinfo {author} {\bibfnamefont {C.}~\bibnamefont {Wang}}, \bibinfo {author}
  {\bibfnamefont {X.}~\bibnamefont {Wang}}, \bibinfo {author} {\bibfnamefont
  {S.~H.}\ \bibnamefont {Wong}}, \bibinfo {author} {\bibfnamefont
  {J.}~\bibnamefont {Wu}}, \bibinfo {author} {\bibfnamefont {X.}~\bibnamefont
  {Wu}}, \bibinfo {author} {\bibfnamefont {X.}~\bibnamefont {Xia}}, \bibinfo
  {author} {\bibfnamefont {Y.}~\bibnamefont {Yan}}, \bibinfo {author}
  {\bibfnamefont {R.~W.-Y.}\ \bibnamefont {Yeung}}, \bibinfo {author}
  {\bibfnamefont {T.~C.}\ \bibnamefont {Yiu}}, \bibinfo {author} {\bibfnamefont
  {S.}~\bibnamefont {Zhang}}, \bibinfo {author} {\bibfnamefont
  {W.}~\bibnamefont {Zhang}}, \bibinfo {author} {\bibfnamefont
  {X.}~\bibnamefont {Zhang}}, \bibinfo {author} {\bibfnamefont
  {Q.}~\bibnamefont {Zhao}}, \ and\ \bibinfo {author} {\bibfnamefont {S.-G.}\
  \bibnamefont {Zhou}},\ }\href {\doibase 10.1016/j.adt.2022.101488} {\bibfield
   {journal} {\bibinfo  {journal} {At. Data Nucl. Data Tables}\ }\textbf
  {\bibinfo {volume} {144}},\ \bibinfo {pages} {101488} (\bibinfo {year}
  {2022})}\BibitemShut {NoStop}%
\bibitem [{\citenamefont {Stoitsov}\ \emph {et~al.}(2003)\citenamefont
  {Stoitsov}, \citenamefont {Dobaczewski}, \citenamefont {Nazarewicz},
  \citenamefont {Pittel},\ and\ \citenamefont
  {Dean}}]{Stoitsov2003_PRC68-054312}%
  \BibitemOpen
  \bibfield  {author} {\bibinfo {author} {\bibfnamefont {M.~V.}\ \bibnamefont
  {Stoitsov}}, \bibinfo {author} {\bibfnamefont {J.}~\bibnamefont
  {Dobaczewski}}, \bibinfo {author} {\bibfnamefont {W.}~\bibnamefont
  {Nazarewicz}}, \bibinfo {author} {\bibfnamefont {S.}~\bibnamefont {Pittel}},
  \ and\ \bibinfo {author} {\bibfnamefont {D.~J.}\ \bibnamefont {Dean}},\
  }\href {\doibase 10.1103/PhysRevC.68.054312} {\bibfield  {journal} {\bibinfo
  {journal} {Phys. Rev. C}\ }\textbf {\bibinfo {volume} {68}},\ \bibinfo
  {pages} {054312} (\bibinfo {year} {2003})}\BibitemShut {NoStop}%
\bibitem [{\citenamefont {Goriely}\ \emph {et~al.}(2009)\citenamefont
  {Goriely}, \citenamefont {Hilaire}, \citenamefont {Girod},\ and\
  \citenamefont {P\'eru}}]{Goriely2009_PRL102-242501}%
  \BibitemOpen
  \bibfield  {author} {\bibinfo {author} {\bibfnamefont {S.}~\bibnamefont
  {Goriely}}, \bibinfo {author} {\bibfnamefont {S.}~\bibnamefont {Hilaire}},
  \bibinfo {author} {\bibfnamefont {M.}~\bibnamefont {Girod}}, \ and\ \bibinfo
  {author} {\bibfnamefont {S.}~\bibnamefont {P\'eru}},\ }\href {\doibase
  10.1103/PhysRevLett.102.242501} {\bibfield  {journal} {\bibinfo  {journal}
  {Phys. Rev. Lett.}\ }\textbf {\bibinfo {volume} {102}},\ \bibinfo {pages}
  {242501} (\bibinfo {year} {2009})}\BibitemShut {NoStop}%
\bibitem [{\citenamefont {Goriely}\ \emph {et~al.}(2010)\citenamefont
  {Goriely}, \citenamefont {Chamel},\ and\ \citenamefont
  {Pearson}}]{Goriely2010_PRC82-035804}%
  \BibitemOpen
  \bibfield  {author} {\bibinfo {author} {\bibfnamefont {S.}~\bibnamefont
  {Goriely}}, \bibinfo {author} {\bibfnamefont {N.}~\bibnamefont {Chamel}}, \
  and\ \bibinfo {author} {\bibfnamefont {J.~M.}\ \bibnamefont {Pearson}},\
  }\href {\doibase 10.1103/PhysRevC.82.035804} {\bibfield  {journal} {\bibinfo
  {journal} {Phys. Rev. C}\ }\textbf {\bibinfo {volume} {82}},\ \bibinfo
  {pages} {035804} (\bibinfo {year} {2010})}\BibitemShut {NoStop}%
\bibitem [{\citenamefont {Piekarewicz}\ \emph {et~al.}(2010)\citenamefont
  {Piekarewicz}, \citenamefont {Centelles}, \citenamefont {Roca-Maza},\ and\
  \citenamefont {Vi\~{n}as}}]{Piekarewicz2010_EPJA46-379}%
  \BibitemOpen
  \bibfield  {author} {\bibinfo {author} {\bibfnamefont {J.}~\bibnamefont
  {Piekarewicz}}, \bibinfo {author} {\bibfnamefont {M.}~\bibnamefont
  {Centelles}}, \bibinfo {author} {\bibfnamefont {X.}~\bibnamefont
  {Roca-Maza}}, \ and\ \bibinfo {author} {\bibfnamefont {X.}~\bibnamefont
  {Vi\~{n}as}},\ }\href {\doibase 10.1140/epja/i2010-11051-8} {\bibfield
  {journal} {\bibinfo  {journal} {Eur. Phys. J. A}\ }\textbf {\bibinfo {volume}
  {46}},\ \bibinfo {pages} {379} (\bibinfo {year} {2010})}\BibitemShut
  {NoStop}%
\bibitem [{\citenamefont {Sun}\ \emph {et~al.}(2014)\citenamefont {Sun},
  \citenamefont {Lu}, \citenamefont {Peng}, \citenamefont {Liu},\ and\
  \citenamefont {Zhao}}]{Sun2014_PRC90-054318}%
  \BibitemOpen
  \bibfield  {author} {\bibinfo {author} {\bibfnamefont {B.~H.}\ \bibnamefont
  {Sun}}, \bibinfo {author} {\bibfnamefont {Y.}~\bibnamefont {Lu}}, \bibinfo
  {author} {\bibfnamefont {J.~P.}\ \bibnamefont {Peng}}, \bibinfo {author}
  {\bibfnamefont {C.~Y.}\ \bibnamefont {Liu}}, \ and\ \bibinfo {author}
  {\bibfnamefont {Y.~M.}\ \bibnamefont {Zhao}},\ }\href {\doibase
  10.1103/PhysRevC.90.054318} {\bibfield  {journal} {\bibinfo  {journal} {Phys.
  Rev. C}\ }\textbf {\bibinfo {volume} {90}},\ \bibinfo {pages} {054318}
  (\bibinfo {year} {2014})}\BibitemShut {NoStop}%
\bibitem [{\citenamefont {Bao}\ \emph {et~al.}(2016)\citenamefont {Bao},
  \citenamefont {Lu}, \citenamefont {Zhao},\ and\ \citenamefont
  {Arima}}]{Bao2016_PRC94-064315}%
  \BibitemOpen
  \bibfield  {author} {\bibinfo {author} {\bibfnamefont {M.}~\bibnamefont
  {Bao}}, \bibinfo {author} {\bibfnamefont {Y.}~\bibnamefont {Lu}}, \bibinfo
  {author} {\bibfnamefont {Y.~M.}\ \bibnamefont {Zhao}}, \ and\ \bibinfo
  {author} {\bibfnamefont {A.}~\bibnamefont {Arima}},\ }\href {\doibase
  10.1103/PhysRevC.94.064315} {\bibfield  {journal} {\bibinfo  {journal} {Phys.
  Rev. C}\ }\textbf {\bibinfo {volume} {94}},\ \bibinfo {pages} {064315}
  (\bibinfo {year} {2016})}\BibitemShut {NoStop}%
\bibitem [{\citenamefont {Sun}\ \emph {et~al.}(2017)\citenamefont {Sun},
  \citenamefont {Liu},\ and\ \citenamefont {Wang}}]{Sun2017_PRC95-014307}%
  \BibitemOpen
  \bibfield  {author} {\bibinfo {author} {\bibfnamefont {B.~H.}\ \bibnamefont
  {Sun}}, \bibinfo {author} {\bibfnamefont {C.~Y.}\ \bibnamefont {Liu}}, \ and\
  \bibinfo {author} {\bibfnamefont {H.~X.}\ \bibnamefont {Wang}},\ }\href
  {\doibase 10.1103/PhysRevC.95.014307} {\bibfield  {journal} {\bibinfo
  {journal} {Phys. Rev. C}\ }\textbf {\bibinfo {volume} {95}},\ \bibinfo
  {pages} {014307} (\bibinfo {year} {2017})}\BibitemShut {NoStop}%
\bibitem [{\citenamefont {Bao}\ \emph {et~al.}(2020)\citenamefont {Bao},
  \citenamefont {Zong}, \citenamefont {Zhao},\ and\ \citenamefont
  {Arima}}]{Bao2020_PRC102-014306}%
  \BibitemOpen
  \bibfield  {author} {\bibinfo {author} {\bibfnamefont {M.}~\bibnamefont
  {Bao}}, \bibinfo {author} {\bibfnamefont {Y.~Y.}\ \bibnamefont {Zong}},
  \bibinfo {author} {\bibfnamefont {Y.~M.}\ \bibnamefont {Zhao}}, \ and\
  \bibinfo {author} {\bibfnamefont {A.}~\bibnamefont {Arima}},\ }\href
  {\doibase 10.1103/PhysRevC.102.014306} {\bibfield  {journal} {\bibinfo
  {journal} {Phys. Rev. C}\ }\textbf {\bibinfo {volume} {102}},\ \bibinfo
  {pages} {014306} (\bibinfo {year} {2020})}\BibitemShut {NoStop}%
\bibitem [{\citenamefont {Ma}\ \emph {et~al.}(2021)\citenamefont {Ma},
  \citenamefont {Zong}, \citenamefont {Zhao},\ and\ \citenamefont
  {Arima}}]{Ma2021_PRC104-014303}%
  \BibitemOpen
  \bibfield  {author} {\bibinfo {author} {\bibfnamefont {C.}~\bibnamefont
  {Ma}}, \bibinfo {author} {\bibfnamefont {Y.~Y.}\ \bibnamefont {Zong}},
  \bibinfo {author} {\bibfnamefont {Y.~M.}\ \bibnamefont {Zhao}}, \ and\
  \bibinfo {author} {\bibfnamefont {A.}~\bibnamefont {Arima}},\ }\href
  {\doibase 10.1103/PhysRevC.104.014303} {\bibfield  {journal} {\bibinfo
  {journal} {Phys. Rev. C}\ }\textbf {\bibinfo {volume} {104}},\ \bibinfo
  {pages} {014303} (\bibinfo {year} {2021})}\BibitemShut {NoStop}%
\bibitem [{\citenamefont {Forss\'en}\ \emph {et~al.}(2009)\citenamefont
  {Forss\'en}, \citenamefont {Caurier},\ and\ \citenamefont
  {Navr\'atil}}]{Forssen2009_PRC79-021303}%
  \BibitemOpen
  \bibfield  {author} {\bibinfo {author} {\bibfnamefont {C.}~\bibnamefont
  {Forss\'en}}, \bibinfo {author} {\bibfnamefont {E.}~\bibnamefont {Caurier}},
  \ and\ \bibinfo {author} {\bibfnamefont {P.}~\bibnamefont {Navr\'atil}},\
  }\href {\doibase 10.1103/PhysRevC.79.021303} {\bibfield  {journal} {\bibinfo
  {journal} {Phys. Rev. C}\ }\textbf {\bibinfo {volume} {79}},\ \bibinfo
  {pages} {021303(R)} (\bibinfo {year} {2009})}\BibitemShut {NoStop}%
\bibitem [{\citenamefont {Akkoyun}\ \emph {et~al.}(2013)\citenamefont
  {Akkoyun}, \citenamefont {Bayram}, \citenamefont {Kara},\ and\ \citenamefont
  {Sinan}}]{Akkoyun2013_JPG40-055106}%
  \BibitemOpen
  \bibfield  {author} {\bibinfo {author} {\bibfnamefont {S.}~\bibnamefont
  {Akkoyun}}, \bibinfo {author} {\bibfnamefont {T.}~\bibnamefont {Bayram}},
  \bibinfo {author} {\bibfnamefont {S.~O.}\ \bibnamefont {Kara}}, \ and\
  \bibinfo {author} {\bibfnamefont {A.}~\bibnamefont {Sinan}},\ }\href
  {\doibase 10.1088/0954-3899/40/5/055106} {\bibfield  {journal} {\bibinfo
  {journal} {J. Phys. G: Nucl. Part. Phys.}\ }\textbf {\bibinfo {volume}
  {40}},\ \bibinfo {pages} {055106} (\bibinfo {year} {2013})}\BibitemShut
  {NoStop}%
\bibitem [{\citenamefont {Wu}\ \emph {et~al.}(2020)\citenamefont {Wu},
  \citenamefont {Bai}, \citenamefont {Sagawa},\ and\ \citenamefont
  {Zhang}}]{Wu2020_PRC102-054323}%
  \BibitemOpen
  \bibfield  {author} {\bibinfo {author} {\bibfnamefont {D.}~\bibnamefont
  {Wu}}, \bibinfo {author} {\bibfnamefont {C.~L.}\ \bibnamefont {Bai}},
  \bibinfo {author} {\bibfnamefont {H.}~\bibnamefont {Sagawa}}, \ and\ \bibinfo
  {author} {\bibfnamefont {H.~Q.}\ \bibnamefont {Zhang}},\ }\href {\doibase
  10.1103/PhysRevC.102.054323} {\bibfield  {journal} {\bibinfo  {journal}
  {Phys. Rev. C}\ }\textbf {\bibinfo {volume} {102}},\ \bibinfo {pages}
  {054323} (\bibinfo {year} {2020})}\BibitemShut {NoStop}%
\bibitem [{\citenamefont {Utama}\ \emph {et~al.}(2016)\citenamefont {Utama},
  \citenamefont {Chen},\ and\ \citenamefont
  {Piekarewicz}}]{Utama2016_JPG43-114002}%
  \BibitemOpen
  \bibfield  {author} {\bibinfo {author} {\bibfnamefont {R.}~\bibnamefont
  {Utama}}, \bibinfo {author} {\bibfnamefont {W.-C.}\ \bibnamefont {Chen}}, \
  and\ \bibinfo {author} {\bibfnamefont {J.}~\bibnamefont {Piekarewicz}},\
  }\href {\doibase 10.1088/0954-3899/43/11/114002} {\bibfield  {journal}
  {\bibinfo  {journal} {J. Phys. G: Nucl. Part. Phys.}\ }\textbf {\bibinfo
  {volume} {43}},\ \bibinfo {pages} {114002} (\bibinfo {year}
  {2016})}\BibitemShut {NoStop}%
\bibitem [{\citenamefont {Neufcourt}\ \emph {et~al.}(2018)\citenamefont
  {Neufcourt}, \citenamefont {Cao}, \citenamefont {Nazarewicz},\ and\
  \citenamefont {Viens}}]{Neufcourt2018_PRC98-034318}%
  \BibitemOpen
  \bibfield  {author} {\bibinfo {author} {\bibfnamefont {L.}~\bibnamefont
  {Neufcourt}}, \bibinfo {author} {\bibfnamefont {Y.}~\bibnamefont {Cao}},
  \bibinfo {author} {\bibfnamefont {W.}~\bibnamefont {Nazarewicz}}, \ and\
  \bibinfo {author} {\bibfnamefont {F.}~\bibnamefont {Viens}},\ }\href
  {\doibase 10.1103/PhysRevC.98.034318} {\bibfield  {journal} {\bibinfo
  {journal} {Phys. Rev. C}\ }\textbf {\bibinfo {volume} {98}},\ \bibinfo
  {pages} {034318} (\bibinfo {year} {2018})}\BibitemShut {NoStop}%
\bibitem [{\citenamefont {Ma}\ \emph {et~al.}(2020)\citenamefont {Ma},
  \citenamefont {Su}, \citenamefont {Liu}, \citenamefont {Ren}, \citenamefont
  {Xu},\ and\ \citenamefont {Gao}}]{Ma2020_PRC101-014304}%
  \BibitemOpen
  \bibfield  {author} {\bibinfo {author} {\bibfnamefont {Y.}~\bibnamefont
  {Ma}}, \bibinfo {author} {\bibfnamefont {C.}~\bibnamefont {Su}}, \bibinfo
  {author} {\bibfnamefont {J.}~\bibnamefont {Liu}}, \bibinfo {author}
  {\bibfnamefont {Z.}~\bibnamefont {Ren}}, \bibinfo {author} {\bibfnamefont
  {C.}~\bibnamefont {Xu}}, \ and\ \bibinfo {author} {\bibfnamefont
  {Y.}~\bibnamefont {Gao}},\ }\href {\doibase 10.1103/PhysRevC.101.014304}
  {\bibfield  {journal} {\bibinfo  {journal} {Phys. Rev. C}\ }\textbf {\bibinfo
  {volume} {101}},\ \bibinfo {pages} {014304} (\bibinfo {year}
  {2020})}\BibitemShut {NoStop}%
\bibitem [{\citenamefont {Dong}\ \emph {et~al.}(2022)\citenamefont {Dong},
  \citenamefont {An}, \citenamefont {Lu},\ and\ \citenamefont
  {Geng}}]{Dong2022_PRC105-014308}%
  \BibitemOpen
  \bibfield  {author} {\bibinfo {author} {\bibfnamefont {X.-X.}\ \bibnamefont
  {Dong}}, \bibinfo {author} {\bibfnamefont {R.}~\bibnamefont {An}}, \bibinfo
  {author} {\bibfnamefont {J.-X.}\ \bibnamefont {Lu}}, \ and\ \bibinfo {author}
  {\bibfnamefont {L.-S.}\ \bibnamefont {Geng}},\ }\href {\doibase
  10.1103/PhysRevC.105.014308} {\bibfield  {journal} {\bibinfo  {journal}
  {Phys. Rev. C}\ }\textbf {\bibinfo {volume} {105}},\ \bibinfo {pages}
  {014308} (\bibinfo {year} {2022})}\BibitemShut {NoStop}%
\bibitem [{\citenamefont {Marquardt}(1963)}]{Marquardt1963_JSIAM11-431}%
  \BibitemOpen
  \bibfield  {author} {\bibinfo {author} {\bibfnamefont {D.~W.}\ \bibnamefont
  {Marquardt}},\ }\href {\doibase 10.1137/0111030} {\bibfield  {journal}
  {\bibinfo  {journal} {J. Soc. Indust. Appl. Math.}\ }\textbf {\bibinfo
  {volume} {11}},\ \bibinfo {pages} {431} (\bibinfo {year} {1963})}\BibitemShut
  {NoStop}%
\bibitem [{\citenamefont {Wang}\ and\ \citenamefont
  {Liu}(2011)}]{Wang2011_PRC84-051303R}%
  \BibitemOpen
  \bibfield  {author} {\bibinfo {author} {\bibfnamefont {N.}~\bibnamefont
  {Wang}}\ and\ \bibinfo {author} {\bibfnamefont {M.}~\bibnamefont {Liu}},\
  }\href {\doibase 10.1103/PhysRevC.84.051303} {\bibfield  {journal} {\bibinfo
  {journal} {Phys. Rev. C}\ }\textbf {\bibinfo {volume} {84}},\ \bibinfo
  {pages} {051303(R)} (\bibinfo {year} {2011})}\BibitemShut {NoStop}%
\bibitem [{\citenamefont {Niu}\ \emph {et~al.}(2013)\citenamefont {Niu},
  \citenamefont {Zhu}, \citenamefont {Niu}, \citenamefont {Sun}, \citenamefont
  {Heng},\ and\ \citenamefont {Guo}}]{Niu2013_PRC88-024325}%
  \BibitemOpen
  \bibfield  {author} {\bibinfo {author} {\bibfnamefont {Z.~M.}\ \bibnamefont
  {Niu}}, \bibinfo {author} {\bibfnamefont {Z.~L.}\ \bibnamefont {Zhu}},
  \bibinfo {author} {\bibfnamefont {Y.~F.}\ \bibnamefont {Niu}}, \bibinfo
  {author} {\bibfnamefont {B.~H.}\ \bibnamefont {Sun}}, \bibinfo {author}
  {\bibfnamefont {T.~H.}\ \bibnamefont {Heng}}, \ and\ \bibinfo {author}
  {\bibfnamefont {J.~Y.}\ \bibnamefont {Guo}},\ }\href {\doibase
  10.1103/PhysRevC.88.024325} {\bibfield  {journal} {\bibinfo  {journal} {Phys.
  Rev. C}\ }\textbf {\bibinfo {volume} {88}},\ \bibinfo {pages} {024325}
  (\bibinfo {year} {2013})}\BibitemShut {NoStop}%
\bibitem [{\citenamefont {Zheng}\ \emph {et~al.}(2014)\citenamefont {Zheng},
  \citenamefont {Wang}, \citenamefont {Wang}, \citenamefont {Niu},
  \citenamefont {Niu},\ and\ \citenamefont {Sun}}]{Zheng2014_PRC90-014303}%
  \BibitemOpen
  \bibfield  {author} {\bibinfo {author} {\bibfnamefont {J.~S.}\ \bibnamefont
  {Zheng}}, \bibinfo {author} {\bibfnamefont {N.~Y.}\ \bibnamefont {Wang}},
  \bibinfo {author} {\bibfnamefont {Z.~Y.}\ \bibnamefont {Wang}}, \bibinfo
  {author} {\bibfnamefont {Z.~M.}\ \bibnamefont {Niu}}, \bibinfo {author}
  {\bibfnamefont {Y.~F.}\ \bibnamefont {Niu}}, \ and\ \bibinfo {author}
  {\bibfnamefont {B.}~\bibnamefont {Sun}},\ }\href {\doibase
  10.1103/PhysRevC.90.014303} {\bibfield  {journal} {\bibinfo  {journal} {Phys.
  Rev. C}\ }\textbf {\bibinfo {volume} {90}},\ \bibinfo {pages} {014303}
  (\bibinfo {year} {2014})}\BibitemShut {NoStop}%
\bibitem [{\citenamefont {Niu}\ \emph {et~al.}(2016)\citenamefont {Niu},
  \citenamefont {Sun}, \citenamefont {Liang}, \citenamefont {Niu},\ and\
  \citenamefont {Guo}}]{Niu2016_PRC94-054315}%
  \BibitemOpen
  \bibfield  {author} {\bibinfo {author} {\bibfnamefont {Z.~M.}\ \bibnamefont
  {Niu}}, \bibinfo {author} {\bibfnamefont {B.~H.}\ \bibnamefont {Sun}},
  \bibinfo {author} {\bibfnamefont {H.~Z.}\ \bibnamefont {Liang}}, \bibinfo
  {author} {\bibfnamefont {Y.~F.}\ \bibnamefont {Niu}}, \ and\ \bibinfo
  {author} {\bibfnamefont {J.~Y.}\ \bibnamefont {Guo}},\ }\href {\doibase
  10.1103/PhysRevC.94.054315} {\bibfield  {journal} {\bibinfo  {journal} {Phys.
  Rev. C}\ }\textbf {\bibinfo {volume} {94}},\ \bibinfo {pages} {054315}
  (\bibinfo {year} {2016})}\BibitemShut {NoStop}%
\bibitem [{\citenamefont {Niu}\ \emph {et~al.}(2018)\citenamefont {Niu},
  \citenamefont {Liang}, \citenamefont {Sun}, \citenamefont {Niu},
  \citenamefont {Guo},\ and\ \citenamefont {Meng}}]{Niu2018_SciBull63-759}%
  \BibitemOpen
  \bibfield  {author} {\bibinfo {author} {\bibfnamefont {Z.}~\bibnamefont
  {Niu}}, \bibinfo {author} {\bibfnamefont {H.}~\bibnamefont {Liang}}, \bibinfo
  {author} {\bibfnamefont {B.}~\bibnamefont {Sun}}, \bibinfo {author}
  {\bibfnamefont {Y.}~\bibnamefont {Niu}}, \bibinfo {author} {\bibfnamefont
  {J.}~\bibnamefont {Guo}}, \ and\ \bibinfo {author} {\bibfnamefont
  {J.}~\bibnamefont {Meng}},\ }\href {\doibase 10.1016/j.scib.2018.05.009}
  {\bibfield  {journal} {\bibinfo  {journal} {Sci. Bull.}\ }\textbf {\bibinfo
  {volume} {63}},\ \bibinfo {pages} {759} (\bibinfo {year} {2018})}\BibitemShut
  {NoStop}%
\bibitem [{\citenamefont {Niu}\ and\ \citenamefont
  {Liang}(2018)}]{Niu2018_PLB778-48}%
  \BibitemOpen
  \bibfield  {author} {\bibinfo {author} {\bibfnamefont {Z.~M.}\ \bibnamefont
  {Niu}}\ and\ \bibinfo {author} {\bibfnamefont {H.~Z.}\ \bibnamefont
  {Liang}},\ }\href {\doibase 10.1016/j.physletb.2018.01.002} {\bibfield
  {journal} {\bibinfo  {journal} {Phys. Lett. B}\ }\textbf {\bibinfo {volume}
  {778}},\ \bibinfo {pages} {48} (\bibinfo {year} {2018})}\BibitemShut
  {NoStop}%
\bibitem [{\citenamefont {Shi}\ \emph {et~al.}(2021)\citenamefont {Shi},
  \citenamefont {Fang},\ and\ \citenamefont
  {Niu}}]{Shi2021_ChinPhysC45-044103}%
  \BibitemOpen
  \bibfield  {author} {\bibinfo {author} {\bibfnamefont {M.}~\bibnamefont
  {Shi}}, \bibinfo {author} {\bibfnamefont {J.~Y.}\ \bibnamefont {Fang}}, \
  and\ \bibinfo {author} {\bibfnamefont {Z.~M.}\ \bibnamefont {Niu}},\ }\href
  {\doibase 10.1088/1674-1137/abdf42} {\bibfield  {journal} {\bibinfo
  {journal} {Chin. Phys. C}\ }\textbf {\bibinfo {volume} {45}},\ \bibinfo
  {pages} {044103} (\bibinfo {year} {2021})}\BibitemShut {NoStop}%
\bibitem [{\citenamefont {Wu}\ and\ \citenamefont
  {Zhao}(2020)}]{Wu2020_PRC101-051301}%
  \BibitemOpen
  \bibfield  {author} {\bibinfo {author} {\bibfnamefont {X.~H.}\ \bibnamefont
  {Wu}}\ and\ \bibinfo {author} {\bibfnamefont {P.~W.}\ \bibnamefont {Zhao}},\
  }\href {\doibase 10.1103/PhysRevC.101.051301} {\bibfield  {journal} {\bibinfo
   {journal} {Phys. Rev. C}\ }\textbf {\bibinfo {volume} {101}},\ \bibinfo
  {pages} {051301(R)} (\bibinfo {year} {2020})}\BibitemShut {NoStop}%
\bibitem [{\citenamefont {Wu}\ \emph {et~al.}(2021)\citenamefont {Wu},
  \citenamefont {Guo},\ and\ \citenamefont {Zhao}}]{Wu2021_PLB819-136387}%
  \BibitemOpen
  \bibfield  {author} {\bibinfo {author} {\bibfnamefont {X.~H.}\ \bibnamefont
  {Wu}}, \bibinfo {author} {\bibfnamefont {L.~H.}\ \bibnamefont {Guo}}, \ and\
  \bibinfo {author} {\bibfnamefont {P.~W.}\ \bibnamefont {Zhao}},\ }\href
  {\doibase 10.1016/j.physletb.2021.136387} {\bibfield  {journal} {\bibinfo
  {journal} {Phys. Lett. B}\ }\textbf {\bibinfo {volume} {819}},\ \bibinfo
  {pages} {136387} (\bibinfo {year} {2021})}\BibitemShut {NoStop}%
\bibitem [{\citenamefont {Wu}\ \emph {et~al.}(2022)\citenamefont {Wu},
  \citenamefont {Ren},\ and\ \citenamefont {Zhao}}]{Wu2022_PRC105-L031303}%
  \BibitemOpen
  \bibfield  {author} {\bibinfo {author} {\bibfnamefont {X.~H.}\ \bibnamefont
  {Wu}}, \bibinfo {author} {\bibfnamefont {Z.~X.}\ \bibnamefont {Ren}}, \ and\
  \bibinfo {author} {\bibfnamefont {P.~W.}\ \bibnamefont {Zhao}},\ }\href
  {\doibase 10.1103/PhysRevC.105.L031303} {\bibfield  {journal} {\bibinfo
  {journal} {Phys. Rev. C}\ }\textbf {\bibinfo {volume} {105}},\ \bibinfo
  {pages} {L031303} (\bibinfo {year} {2022})}\BibitemShut {NoStop}%
\bibitem [{\citenamefont {Angeli}\ \emph {et~al.}(2009)\citenamefont {Angeli},
  \citenamefont {Gangrsky}, \citenamefont {Marinova}, \citenamefont {Boboshin},
  \citenamefont {Komarov}, \citenamefont {Ishkhanov},\ and\ \citenamefont
  {Varlamov}}]{Angeli2009_JPG36-085102}%
  \BibitemOpen
  \bibfield  {author} {\bibinfo {author} {\bibfnamefont {I.}~\bibnamefont
  {Angeli}}, \bibinfo {author} {\bibfnamefont {Y.~P.}\ \bibnamefont
  {Gangrsky}}, \bibinfo {author} {\bibfnamefont {K.~P.}\ \bibnamefont
  {Marinova}}, \bibinfo {author} {\bibfnamefont {I.~N.}\ \bibnamefont
  {Boboshin}}, \bibinfo {author} {\bibfnamefont {S.~Y.}\ \bibnamefont
  {Komarov}}, \bibinfo {author} {\bibfnamefont {B.~S.}\ \bibnamefont
  {Ishkhanov}}, \ and\ \bibinfo {author} {\bibfnamefont {V.~V.}\ \bibnamefont
  {Varlamov}},\ }\href {\doibase 10.1088/0954-3899/36/8/085102} {\bibfield
  {journal} {\bibinfo  {journal} {J. Phys. G: Nucl. Part. Phys.}\ }\textbf
  {\bibinfo {volume} {36}},\ \bibinfo {pages} {085102} (\bibinfo {year}
  {2009})}\BibitemShut {NoStop}%
\bibitem [{\citenamefont {M\"oller}\ \emph {et~al.}(2016)\citenamefont
  {M\"oller}, \citenamefont {Sierk}, \citenamefont {Ichikawa},\ and\
  \citenamefont {Sagawa}}]{Moller2016_ADNDT109-110-1}%
  \BibitemOpen
  \bibfield  {author} {\bibinfo {author} {\bibfnamefont {P.}~\bibnamefont
  {M\"oller}}, \bibinfo {author} {\bibfnamefont {A.~J.}\ \bibnamefont {Sierk}},
  \bibinfo {author} {\bibfnamefont {T.}~\bibnamefont {Ichikawa}}, \ and\
  \bibinfo {author} {\bibfnamefont {H.}~\bibnamefont {Sagawa}},\ }\href
  {\doibase 10.1016/j.adt.2015.10.002} {\bibfield  {journal} {\bibinfo
  {journal} {At. Data Nucl. Data Tables}\ }\textbf {\bibinfo {volume}
  {109-110}},\ \bibinfo {pages} {1} (\bibinfo {year} {2016})}\BibitemShut
  {NoStop}%
\bibitem [{\citenamefont {Li}\ \emph {et~al.}(2021)\citenamefont {Li},
  \citenamefont {Luo},\ and\ \citenamefont {Wang}}]{Li2021_ADNDT140-101440}%
  \BibitemOpen
  \bibfield  {author} {\bibinfo {author} {\bibfnamefont {T.}~\bibnamefont
  {Li}}, \bibinfo {author} {\bibfnamefont {Y.}~\bibnamefont {Luo}}, \ and\
  \bibinfo {author} {\bibfnamefont {N.}~\bibnamefont {Wang}},\ }\href {\doibase
  10.1016/j.adt.2021.101440} {\bibfield  {journal} {\bibinfo  {journal} {At.
  Data Nucl. Data Tables}\ }\textbf {\bibinfo {volume} {140}},\ \bibinfo
  {pages} {101440} (\bibinfo {year} {2021})}\BibitemShut {NoStop}%
\bibitem [{\citenamefont {Kim}\ \emph {et~al.}(2012)\citenamefont {Kim},
  \citenamefont {Jeong}, \citenamefont {Jeong},\ and\ \citenamefont
  {Young}}]{Kim2012_IEEE42-1011}%
  \BibitemOpen
  \bibfield  {author} {\bibinfo {author} {\bibfnamefont {N.}~\bibnamefont
  {Kim}}, \bibinfo {author} {\bibfnamefont {Y.-S.}\ \bibnamefont {Jeong}},
  \bibinfo {author} {\bibfnamefont {M.-K.}\ \bibnamefont {Jeong}}, \ and\
  \bibinfo {author} {\bibfnamefont {T.~M.}\ \bibnamefont {Young}},\ }\href
  {\doibase 10.1109/TSMCC.2011.2177969} {\bibfield  {journal} {\bibinfo
  {journal} {IEEE Trans. Syst. Man Cybern.}\ }\textbf {\bibinfo {volume}
  {42}},\ \bibinfo {pages} {1011} (\bibinfo {year} {2012})}\BibitemShut
  {NoStop}%
\bibitem [{\citenamefont {Wu}\ \emph {et~al.}(2017)\citenamefont {Wu},
  \citenamefont {Fang}, \citenamefont {Chang},\ and\ \citenamefont
  {Kung}}]{Wu2017_IEEE47-3916}%
  \BibitemOpen
  \bibfield  {author} {\bibinfo {author} {\bibfnamefont {P.-Y.}\ \bibnamefont
  {Wu}}, \bibinfo {author} {\bibfnamefont {C.-C.}\ \bibnamefont {Fang}},
  \bibinfo {author} {\bibfnamefont {J.~M.}\ \bibnamefont {Chang}}, \ and\
  \bibinfo {author} {\bibfnamefont {S.-Y.}\ \bibnamefont {Kung}},\ }\href
  {\doibase 10.1109/TCYB.2016.2590472} {\bibfield  {journal} {\bibinfo
  {journal} {IEEE Trans. Cybern}\ }\textbf {\bibinfo {volume} {47}},\ \bibinfo
  {pages} {3916} (\bibinfo {year} {2017})}\BibitemShut {NoStop}%
\bibitem [{\citenamefont {Angeli}(2004)}]{Angeli2004_ADNDT87-185}%
  \BibitemOpen
  \bibfield  {author} {\bibinfo {author} {\bibfnamefont {I.}~\bibnamefont
  {Angeli}},\ }\href {\doibase https://doi.org/10.1016/j.adt.2004.04.002}
  {\bibfield  {journal} {\bibinfo  {journal} {At. Data Nucl. Data Tables}\
  }\textbf {\bibinfo {volume} {87}},\ \bibinfo {pages} {185 } (\bibinfo {year}
  {2004})}\BibitemShut {NoStop}%
\bibitem [{\citenamefont {M\"{o}ller}\ and\ \citenamefont
  {Nix}(1995)}]{Moeller1995_ADNDT59-185}%
  \BibitemOpen
  \bibfield  {author} {\bibinfo {author} {\bibfnamefont {P.}~\bibnamefont
  {M\"{o}ller}}\ and\ \bibinfo {author} {\bibfnamefont {J.}~\bibnamefont
  {Nix}},\ }\href {\doibase 10.1006/adnd.1995.1002} {\bibfield  {journal}
  {\bibinfo  {journal} {At. Data and Nucl. Data Tables}\ }\textbf {\bibinfo
  {volume} {59}},\ \bibinfo {pages} {185} (\bibinfo {year} {1995})}\BibitemShut
  {NoStop}%
\end{thebibliography}

%

\end{multicols}

\end{document}